\documentclass[structabstract]{aa}  
\usepackage{color}
\usepackage{graphicx}
\usepackage{natbib}
\bibpunct{(}{)}{;}{a}{}{,} 
\usepackage{txfonts}
\begin{document}
\title{The CORALIE survey for southern extrasolar planets}

\subtitle{XVII. New and updated long period and massive planets \thanks{The {\footnotesize CORALIE} radial velocity measurements discussed in this paper are only available in electronic form at the CDS via anonymous ftp to cdsarc.u-strasbg.fr (130.79.128.5) or via http://cdsweb.u-strasbg.fr/cgi-bin/qcat?J/A+A/} \thanks{Based on observations collected with the {\footnotesize CORALIE} echelle spectrograph mounted on the 1.2m Swiss telescope at La Silla Observatory and with the HARPS spectrograph on the ESO 3.6 m telescope at La Silla (ESO, Chile)}}

\author{M.~Marmier\inst{1}
         \and D.~S\'egransan\inst{1}
         \and S.~Udry\inst{1}
         \and M.~Mayor\inst{1}
         \and F.~Pepe\inst{1}
         \and D.~Queloz\inst{1}
	 \and C.~Lovis\inst{1}
	 \and D.~Naef\inst{1}
         \and N.C.~Santos\inst{2,3,1}
         \and R.~Alonso\inst{4,5,1}
         \and S.~Alves\inst{8,1}
         \and S.~Berthet\inst{1}
         \and B.~Chazelas\inst{1}
         \and B-O.~Demory\inst{9,1}
         \and X.~Dumusque\inst{1}
         \and A.~Eggenberger\inst{1}
         \and P.~Figueira\inst{2,1} 
         \and M.~Gillon\inst{6,1}  
	 \and J.~Hagelberg\inst{1}   
	 \and M.~Lendl\inst{1}
	 \and R.~A.~Mardling\inst{7,1}
         \and D.~M\'egevand\inst{1}
         \and M.~Neveu\inst{1}
	 \and J.~Sahlmann\inst{1}
         \and D.~Sosnowska\inst{1} 
         \and M.~Tewes\inst{10}
         \and A.~H.M.J.~Triaud\inst{1}
}
\offprints{Maxime Marmier, \email{Maxime.Marmier@unige.ch}}
\institute{Observatoire astronomique de l'Universit\'e de Gen\`eve,
                   51 ch. des Maillettes - Sauverny -, CH-1290 Versoix,
                   Switzerland
   \and
   Centro de Astrof\'{\i}sica, Universidade do Porto, Rua das Estrelas, 4150-762 Porto, Portugal
   \and
   Departamento de F\'{\i}sica e Astronomia, Faculdade de Ci\^encias, Universidade do Porto, Rua do Campo Alegre, 4169-007 Porto, Portugal
   \and
   Instituto de Astrof\'{\i}sica de Canarias, C/ V\'\i a L\'actea S/N, E-38200 La Laguna, Spain
   \and
   Departamento de Astrof\'{\i}sica, Universidad de La Laguna, E-38205 La Laguna, Spain
   \and
   Universit\'e de Li\`ege, All\'ee du 6 ao\^ut 17, Sart Tilman, Li\`ege 1, Belgium
   \and
   School of Mathematical Sciences, Monash University, Victoria, 3800, Australia
   \and
   Departamento de F\'{\i}sica, Universidade Federal do Rio Grande do Norte, 59072-970, Natal, RN., Brazil
   \and
   Department of Earth, Atmospheric and Planetary Sciences, Department of Physics, Massachusetts Institute of Technology, 77 Massachusetts Ave., Cambridge, MA 02139, USA
   \and
   Laboratoire d'astrophysique, Ecole Polytechnique F\'ed\'erale de Lausanne (EPFL), Observatoire de Sauverny, CH-1290 Versoix, Switzerland
}  

\date{Received month day, year; accepted month day, year}

\abstract
{Since 1998, a planet-search program around main sequence stars within 50 pc in the southern hemisphere has been carried out with the CORALIE echelle spectrograph at La Silla Observatory.}
{With an observing time span of more than 14 years, the CORALIE survey is now able to unveil Jovian planets on Jupiter's period domain. This growing period-interval coverage is important for building formation and migration models since observational constraints are still weak for periods beyond the ice line.}
{Long-term precise Doppler measurements with the CORALIE echelle spectrograph, together with a few additional observations made with the HARPS spectrograph on the ESO 3.6\,m telescope, reveal radial velocity signatures of massive planetary companions on long-period orbits.}
{In this paper we present seven new planets orbiting  HD\,27631, HD\,98649, HD\,106515A, HD\,166724, HD\,196067, HD\,219077, and HD\,220689, together with the  CORALIE orbital parameters for three already known planets around HD\,10647, HD\,30562, and HD\,86226. The period range of the new planetary companions goes from 2200 to 5500 days and covers a mass domain between 1 and 10.5 M$_{\mathrm{Jup}}$. Surprisingly, five of them present very high eccentricities above $e>0.57$.  A pumping scenario by Kozai mechanism may be invoked for HD\,106515Ab and HD\,196067b, which are both orbiting stars in multiple systems. Since the presence of a third massive body cannot be inferred from the data of HD\,98649b, HD\,166724b, and HD\,219077b, the origin of the eccentricity of these systems remains unknown. Except for HD\,10647b, no constraint on the upper mass of the planets is provided by Hipparcos astrometric data. Finally, the hosts of these long period planets show no metallicity excess.}
{}


\maketitle
%

\section{Introduction}
\begin{figure}
   \centering
   \begin{minipage}[]{.49\linewidth}
    \begin{center}
       \includegraphics[width=4.5cm]{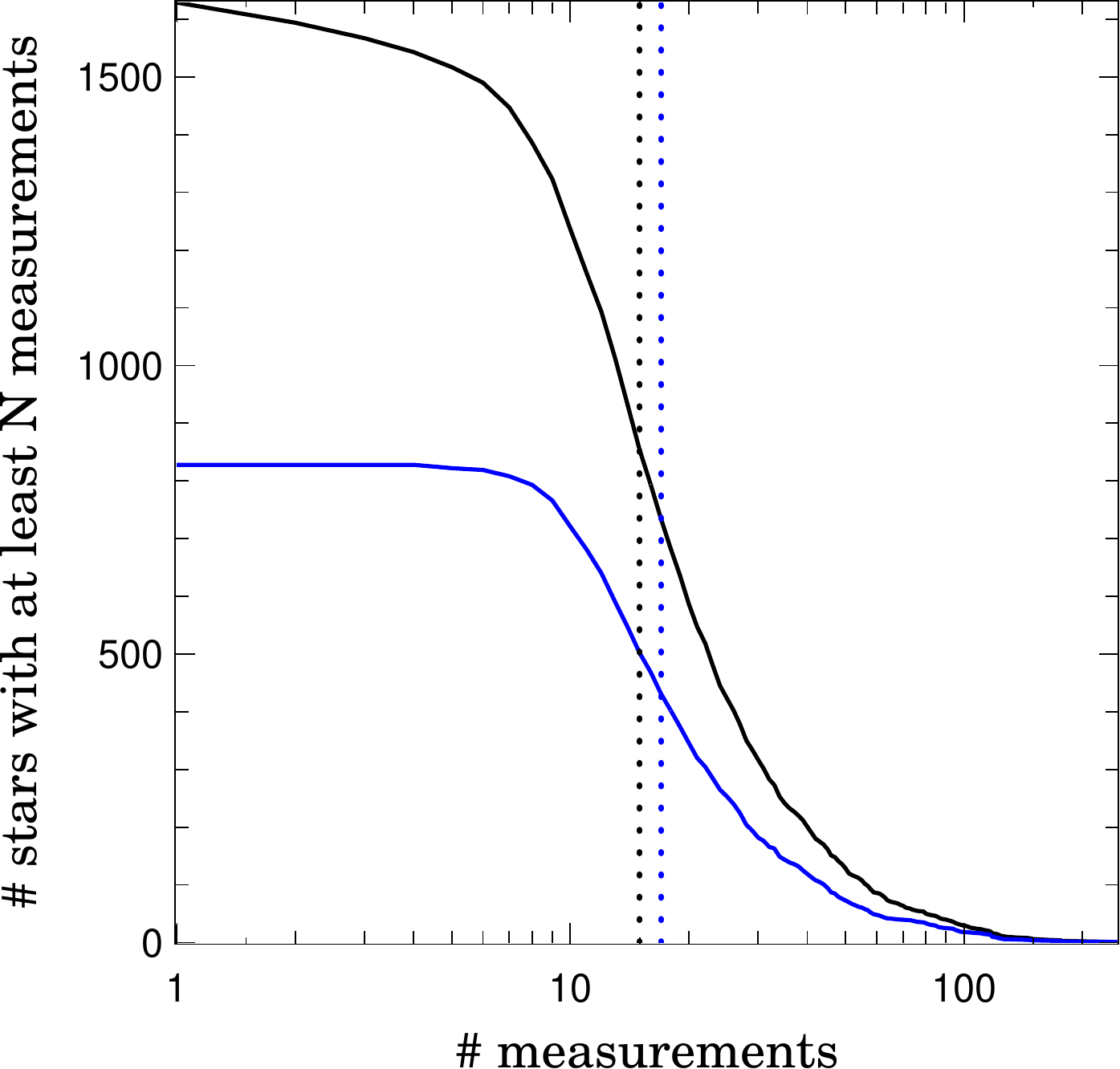}
    \end{center}
   \end{minipage}
   \hfill
   \begin{minipage}[]{.49\linewidth}
    \begin{center}
       \includegraphics[width=4.5cm]{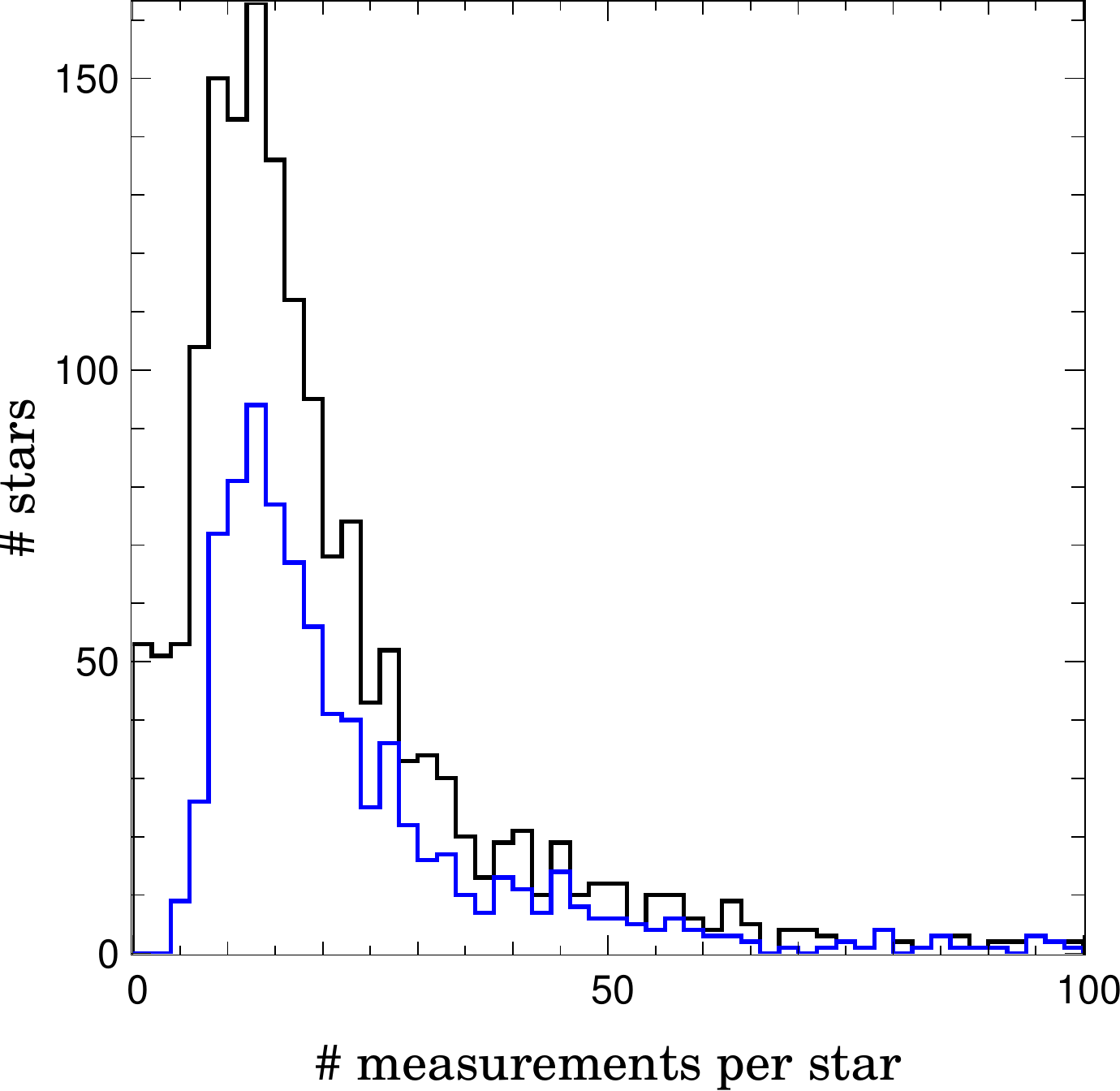}
    \end{center}
   \end{minipage}
   \caption{Statistical distribution of the number of measurements in the CORALIE volume-limited sample (1647 stars). The blue curve represents the subsample (822 stars) of the best candidates for a Doppler-planet search survey (low chromospheric activity index, no high-rotation stars, and no binaries), while the black line is for the entire sample. The histograms of the number of Doppler measurements are plotted on the right, and the curves in the left panel are the inverse cumulative distribution functions of measurements (i.e. the number of stars with at least N data points). The dashed lines represent the median of the distributions for all them and for the subsample (respectively 15 and 17 Doppler measurements).}
   \label{histo_obs_survey}
   \end{figure}

The CORALIE planet-search program in the southern hemisphere \citep{udry2000coralie} has been going on for more than 14 years (since June 1998). This high-precision radial velocity survey makes use of the CORALIE fiber-fed echelle spectrograph mounted on the 1.2 meter Euler Swiss telescope located at La Silla Observatory (ESO, Chile). The stars being monitored are part of a volume-limited sample. It contains 1647 main sequence stars from F8 down to K0 within 50 pc and has a color-dependent distance limit for later type stars down to M0 with the faintest target not exceeding V=10. Today the survey comprises more than 35\,000 radial-velocity measurements on the 1647 stars of the program.\\
If additional constraints are added to reject the binaries, the high-rotation stars and the targets with a chromospheric activity index greater than log\,$R'_{HK} = -4.75$ (or more than $-4.70$ for K stars), half of the original sample remain (822 stars), constituting the best candidates for a Doppler planet search survey. This subsample is observed with a precision of 6 ms$^{-1}$ or better, while the remaining part of the sample suffers from a degraded accuracy due to stellar jitter and/or the widening of the cross-correlation function (induced by high stellar rotation velocities). However, ninety percent of the targets are still monitored with a long-term precision better than 10 ms$^{-1}$. The statistical distribution of number of measurements per star for the entire sample and the subsample (of the best candidates for Doppler-planet search survey) are represented in Fig.\,\ref{histo_obs_survey}. Even if most of the discovered planets were found around stars belonging to the ``quiet" subsample \citep[for more details see][]{mayor2011}, the remaining part of the survey has also been observed well. The median of the distribution for the entire sample and the subsample are respectively 15 and 17 Doppler measurements, while ninety percent of the stars have at least six data points.\\
So far CORALIE has allowed (or has contributed to) detection of more than eighty extrasolar planet candidates. In this paper we report the discovery of seven long-period and massive planets orbiting  HD\,27631, HD\,98649, HD\,106515A, HD\,166724, HD\,196067, HD\,219077, and HD\,220689. We also provide updates for three known systems around HD\,10647 \citep{butler2006}, HD\,30562 \citep{fischer2009longp}, and HD\,86226 \citep{arriagada2009}. After 14 years of activity, the CORALIE survey is now able to unveil planetary candidates with long orbital periods. HD\,106515Ab and HD\,196067b, with periods around ten years, have just completed one full revolution since the beginning of our observations. In the case of the companions orbiting HD\,98649, HD\,166724, and HD\,219077, with periods around five thousand days, the orbital phase coverage is not complete yet. This growing period-interval coverage is very important with regard to formation and migration models since only 28 of the 659 exoplanets (including 5 systems presented in this paper) with known radial velocity or/and astrometric orbits have periods longer than nine years. Interestingly the eccentricities of these five systems are all above 0.57. With eccentricites of 0.85, 0.734, and 0.770, HD\,98649b, HD\,166724b, and HD\,219077b, respectively, are the three planetary companions with the most eccentric orbits on periods longer than five years (Fig.\,\ref{eccvsp}). Finally, with a mass higher than 6M$_{\mathrm{Jup}}$, HD\,98649, HD\,106515A, HD\,196067, and HD\,219077 belong to the upper 15\% of the planetary mass distribution, and they enrich the statistical sample towards the region of the brown-dwarf desert.\\
The paper is organized as follows. The new host stars properties are summarized in Sect.\,\ref{stellar}. Section\,\ref{velocities} presents our radial-velocity data and the inferred orbital solution of the newly detected companions. Section\,\ref{updated_param} presents the stellar properties and CORALIE updated parameters of the three already known planets. These results are discussed in Sect.\,\ref{conclusions} with some concluding remarks.
          
%
\begin{table*}
\caption{Observed and inferred stellar parameters for HD\,27631, HD\,98649, HD\,106515A, HD\,166724, HD\,196067, HD\,219077, and HD\,220689.}
\label{table_stars} 
\tabcolsep=5.6pt     
\centering          
\begin{tabular}{l l c c c c c c c c c}    
\hline\hline       
Parameters		&	 	&HD27631		&HD98649		&HD106515A		&HD166724		&HD196067		&HD219077		&HD220689		\\ 
\hline                    
 Spectral type$^{(a)}$	&		&G3\,IV/V		&G4\,V			&G5    			&K0\,IV/V		&G1\,V			&G8\,V+			&G3\,V			\\  
 $V^{(a)}$		&		&8.26			&8.00			&7.35			&9.33			&6.51			&6.12			&7.74			\\
 $B-V^{(a)}$		&		&0.682			&0.658			&0.815			&0.861			&0.620			&0.787			&0.603			\\
 $\pi^{(b)}$		&[mas]		&22.0$\pm$0.7		&24.1$\pm$0.8		&28.4$\pm$1.0		&23.6$\pm$1.2		&22.6$\pm$2.6		&34.07$\pm$0.37		&22.4$\pm$0.7		\\
 $M_{v}$		&		&4.97			&4.91			&4.62			&6.20			&3.28			&3.78			&4.49			\\
 $T_{eff}$		&[K]		&5737$\pm$36$^{(c)}$	&5759$\pm$35$^{(c)}$	&5362$\pm$29$^{(c)}$	&5127$\pm$52$^{(d)}$	&6017$\pm$46$^{(c)}$	&5362$\pm$18$^{(d)}$	&5921$\pm$26$^{(c)}$	\\
 log\,$g$		&[cgs]		&4.48$\pm$0.09$^{(c)}$	&4.38$\pm$0.08$^{(c)}$	&4.41$\pm$0.04$^{(c)}$	&4.43$\pm$0.08$^{(d)}$	&4.24$\pm$0.03$^{(c)}$	&4.00$\pm$0.03$^{(d)}$	&4.32$\pm$0.03$^{(c)}$	\\
 $[Fe/H]$		&[dex]		&-0.12$\pm$0.05$^{(c)}$	&0.02$\pm$0.03$^{(c)}$	&0.03$\pm$0.02 $^{(c)}$	&-0.09$\pm$0.03$^{(d)}$	&0.18$\pm$0.04$^{(c)}$&-0.13$\pm$0.01$^{(d)}$	&0.00$\pm$0.03$^{(c)}$	\\
 v$sin\,(i)^{(e)}$ 	&[kms$^{-1} $]	&1.34			&1.19			&$<1$			&1.42			&1.45			&1.46			&2.3			\\
 M$_{*}^{(f)}$		&[M$_{\odot}$]	&0.94$\pm$0.04		&1.00$\pm$0.03		&0.97$\pm$0.01		&0.81$\pm$0.02		&1.29$\pm$0.08		&1.05$\pm$0.02	&1.04$\pm$0.03		\\
 $L^{(g)}$		&[L$_{\odot}$]  &0.88			&0.86			&1.23			&0.31			&3.73			&2.66			&1.24			\\   
 $R^{(f)}$		&[R$_{\odot}$] 	&$\approx$1.0		&$\approx$1.0		&1.62$\pm$0.05		&	---		&1.73$\pm$0.21		&1.91$\pm$0.03		&1.07$\pm$0.04		\\   
 log\,$R'_{HK}$		&		&-4.94$\pm$0.08$^{(c)}$	&-4.99$\pm$0.08$^{(c)}$	&-5.09$\pm$0.09$^{(c)}$	&-4.73$\pm$0.03$^{(h)}$	&-5.11$\pm$0.08$^{(c)}$	&-5.12$\pm$0.01$^{(h)}$	&-4.98$\pm$0.05$^{(c)}$	\\
 $P_{rot}^{(i)}$	&[days]		&28$\pm$4		&27$\pm$4		&50$\pm$5		&30$\pm$5		&26$\pm$3		&49$\pm$4		&20$\pm$3		\\
 Age$^{(f)}$		&[Gyr]		&4.4$\pm$3.6		&2.3$\pm$2.0		&11.7$\pm$0.2		&4.0$\pm$3.8		&3.3$\pm$0.6		&8.9$\pm$0.3		&3.5$\pm$1.9		\\     
\hline         
\end{tabular}
\tablefoot{
\tablefoottext{a}{Parameter from HIPPARCOS \citep{esa1997}.}
\tablefoottext{b}{Parallax from the new Hipparcos reduction \citep{vanleeuwen2007}.}
\tablefoottext{c}{Parameter derived using CORALIE spectra.}
\tablefoottext{d}{Parameter derived using HARPS spectra \citep{sousa2008}.}
\tablefoottext{e}{Parameter derived using CORALIE CCF \citep{santos2001feh}.}
\tablefoottext{f}{Parameter derived from \cite{girardi2000} models.}
\tablefoottext{g}{The bolometric correction is computed from \cite{flower1996}.}
\tablefoottext{h}{Parameter derived using HARPS spectra \citep{lovis2011}.}
\tablefoottext{i}{From the calibration of the rotational period vs. activity \citep{mamajek2008}.}
}
\end{table*}	 

\begin{figure}
   \centering
   \includegraphics[width=8.2cm]{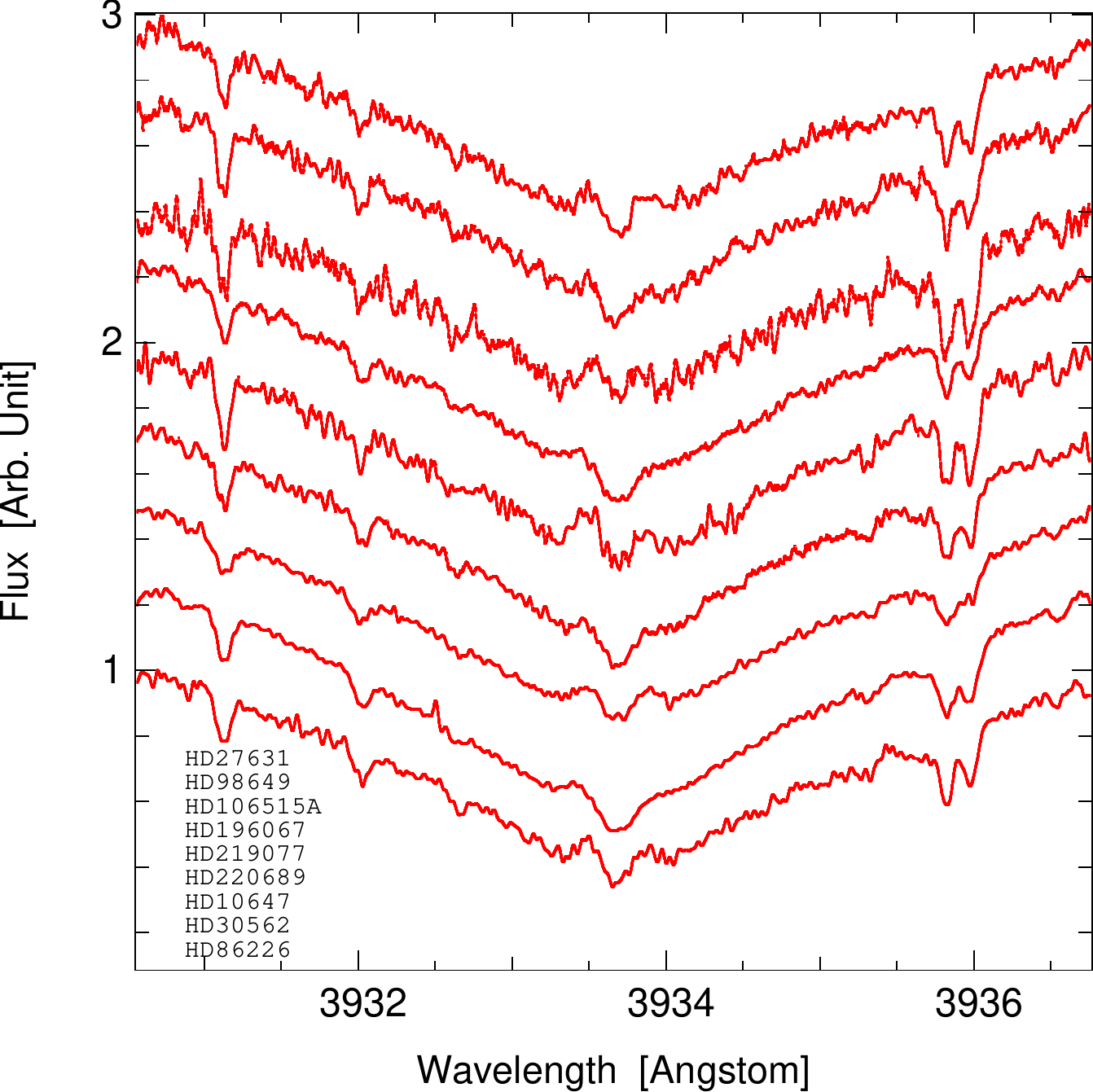}
   \includegraphics[width=8.2cm]{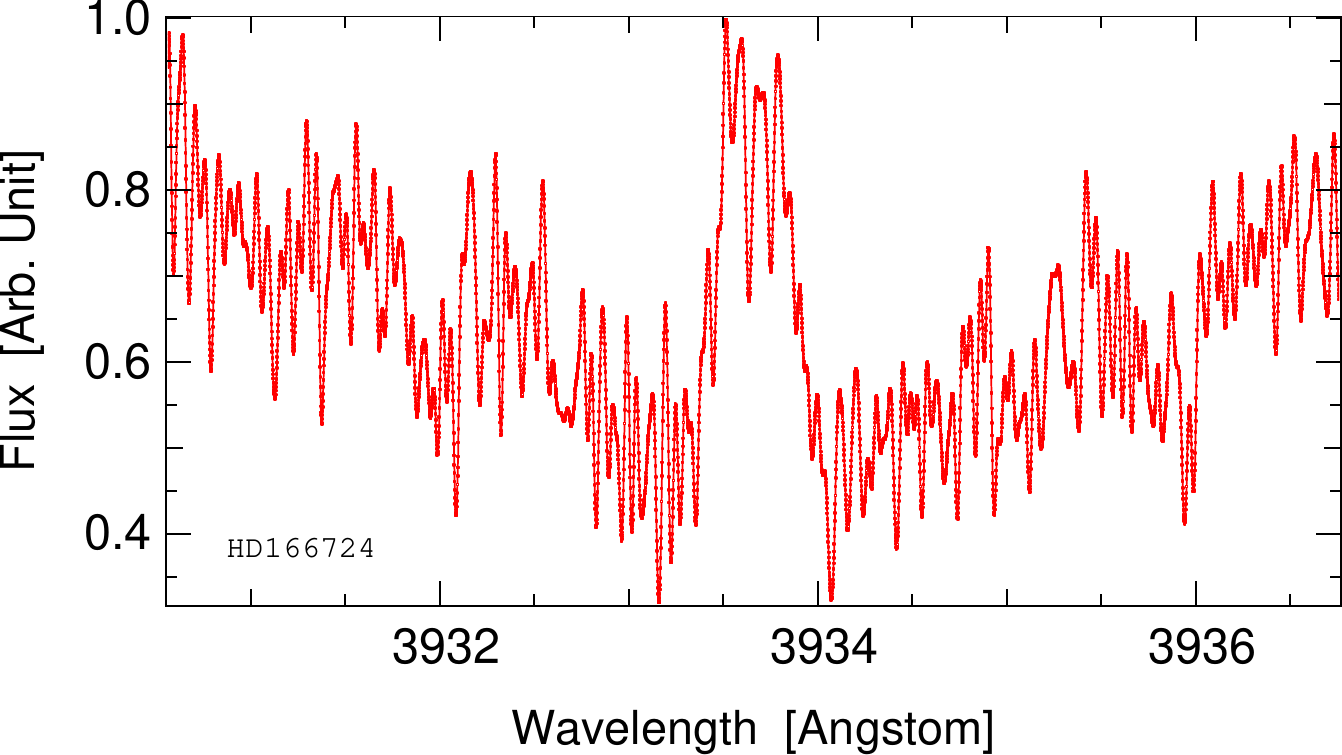}
   \caption{CaII K absorption line region ($\lambda$ = 3933.6 $\AA$) of the co-added CORALIE spectra. This region is contaminated by a few thin thorium emission lines that were filtered while co-adding the spectra. No significant re-emission lines are present at the bottom of the CaII absorption region for the stars displayed in the upper panel, from top to bottom HD\,27631, HD\,98649, HD\,106515A, HD\,196067, HD\,219077, HD\,220689, HD\,10647, HD\,30562, and HD\,86226 . The large re-emission for HD\,166724 (lower panel) indicates the significant chromospheric activity of the star.}
   \label{caII}
\end{figure}

\section{Stellar characteristics}
\label{stellar}
Spectral types and photometric parameters were taken from HIPPARCOS \cite{esa1997}. The astrometric parallaxes come from the improved Hipparcos reduction \citep{vanleeuwen2007}. The atmospheric parameters $T_{eff}$, log $g$, $[Fe/H]$ are from \cite{santos2001feh} and \cite{sousa2008} while the v$sin\,(i)$ was computed using the \cite{santos2002} calibration of CORALIE's cross-correlation function (CCF). Using the stellar activity index log\,$R'_{HK}$, derived from HARPS or CORALIE high signal-to-noise spectra, we derived the rotational period from the empirical calibration described in \cite{mamajek2008}. The bolometric correction was computed from \cite{flower1996} using the spectroscopic $T_{eff}$ determinations. A Bayesian estimation method described in \cite{dasilva2006} and theoretical isochrones from \cite{girardi2000} were used to estimate stellar ages, masses and radii.\footnote{The web interface (PARAM 1.1) for the Bayesian estimation of stellar parameters can be found at http://stev.oapd.inaf.it/cgi-bin/param} The observed and inferred stellar parameters are summarized in Table\,\ref{table_stars}.

\subsection{HD\,27631 (HIP\,20199)}
HD\,27631 is identified as a G3\,IV/V in the Hipparcos catalog \cite{esa1997} with a parallax of $\pi = 22.0\pm0.7$\,mas and an apparent magnitude $V= 8.26$. The spectroscopic analysis of the CORALIE spectra results in an effective temperature of $T_{eff}=5737\pm36$, a stellar metallicity of $[Fe/H]=-0.12\pm0.05$, and a surface gravity log\,$g$ = 4.48$\pm$0.09. Using these parameters and theoretical isochrones from \cite{girardi2000}, we obtain a mass of M$_{*}=0.94\pm0.04$ M$_{\odot}$ with an unconstrained age of $4.4\pm3.6$ Gyr. No emission peak is visible for HD\,27631 at the center of the Ca II K absorption line region (Fig.\,\ref{caII}). We derive a log\,$R'_{HK}$=$-4.94\pm0.08$ compatible with the value of -4.91 found by \cite{henry1996lprhk}.

\subsection{HD\,98649 (HIP\,55409, LTT\,4199)}
HD\,98649 is a chromospherically-quiet G4 dwarf with an astrometric parallax of $\pi = 24.1\pm0.8$\,mas, which sets the star at a distance of $41.5\pm1.4$ pc from the Sun. The apparent magnitude $V= 8.00$ implies an absolute magnitude of $M_{v}=4.91$. According to the Hipparcos catalog, the color index for HD\,98649 is $B$-$V$ = 0.658. Using a bolometric correction $BC=-0.083$ and the solar absolute magnitude $M_{bol}=4.83$ \citep{lejeune1998}, we obtain a luminosity $L=0.86\,L_{\odot}$. Our spectral analysis results in an effective temperature of $T_{eff}=5759\pm35$ and a stellar metallicity of $[Fe/H]=-0.02\pm0.03$. We finally derived a mass of M$_{*}=1.00\pm0.03$ M$_{\odot}$ and an age of $2.3\pm2.0$ Gyr. HD\,98649 has an activity index log\,$R'_{HK}=-4.99\pm0.08$, in agreement with the value published by \cite{jenkins2008lprhk}. The projected rotational velocity is v$sin\,(i)=1.19$ kms$^{-1}$ with a rotational period of $P_{rot}=27\pm4$ days.

\subsection{HD\,106515A (HIP\,59743, CCDM\,J12151-716A, LTT\,4599)}
HD\,106515A is the primary star of a triple visual system located in the Virgo constellation at $35.2\pm1.2$ pc from the Sun. It has spectral type G5, an apparent magnitude $V= 7.35$ with an effective temperature of $T_{eff}=5362\pm29$ and a stellar metallicity of $[Fe/H]=0.03\pm0.02$. The derived mass is M$_{*}=0.97\pm0.01$ M$_{\odot}$ with an age of $11.7\pm0.2$ Gyr. The star is thus a very slow rotator  v$sin\,(i) < 1$ kms$^{-1}$ with a  low activity index log\,$R'_{HK}$=-5.09$\pm$0.09. These values agree closely with the result published by \cite{desidera2006lprhkvsini} (v$sin\,(i) = 0.6$ kms$^{-1}$, log\,$R'_{HK}$=-5.04). HD\,106515A's first companion LTT\,4598 (CCDM\,J12151-716B) is a G8 dwarf 7.5 arcsecond away with an apparent magnitude V=8.3.\\
Our spectroscopic analysis results in $T_{eff}=5324\pm43$, $[Fe/H]=0.09\pm0.03$, log\,$g=4.47\pm0.08$, and a derived mass M$_{*}=0.89$ M$_{\odot}$. Owing to their small angular separation, the two stars have the same name in the Hipparcos catalog. \cite{gould2004} derived an Hipparcos-based parallax for LTT\,4598 and confirm that the two stars have common proper motion, implying that the pair is gravitationally bound. Using the positions given by \cite{gould2004}, we obtain a projected binary separation of 310 AU, which leads to an estimated semi-major axis of 390 AU with the statistical relation $a/r=1.26$ \citep{fischer1992avsr}. A second companion at a separation of 98.7 arcsecond with a magnitude $V= 10.3$,  identified as CCDM\,J12151-716C (BD-06\,3533, TYC\,4946-202-1) is listed in the CCDM catalog \citep{dommanget2002ccdm}.  However, the Tycho-2 catalog \citep{hog2000tycho2} provides different proper motion than for HD\,106515A. The pair is thus probably not physical. \\

\subsection{HD\,166724 (HIP\,89354)}
HD\,166724 has a spectral type K0\,IV/V in the Hipparcos catalog with an apparent V band magnitude $V=9.33$ and a parallax $\pi = 23.6\pm1.2$\,mas. The spectral analysis results in an effective temperature of $T_{eff}=5127\pm52$ and a stellar metallicity of $[Fe/H]=-0.09\pm0.03$. We derived a mass of M$_{*}=0.81\pm0.02$ M$_{\odot}$ with a luminosity of $L=0.31\,L_{\odot}$ and an unconstrained age of $4.0\pm3.8$ Gyr. HD\,166724 is slightly active with a mean stellar activity index derived form the HARPS spectra of log\,$R'_{HK}=-4.73\pm0.03$. A strong re-emission peak is visible for HD\,166724 at the center of the Ca II K absorption line (Fig.\ref{caII}), which indicates the significant chromospherical activity of the star.

\subsection{HD\,196067 (HIP\,102125, CCDM\,J20417-7521A, LTT\,8159)}
HD\,196067 is the primary component of a bright visual binary located at $44\pm5$ pc from the Sun in the Octans constellation. Its spectral type is G1 with an apparent magnitude $V=6.51$. The effective temperature and metallicity are $T_{eff}=6017\pm46$ and $[Fe/H]=0.18\pm0.04$. We derived a mass M$_{*}=1.29\pm0.08$ M$_{\odot}$, a radius of $R= 1.73\pm0.21\,R_{\odot}$, and an age of $3.3\pm0.6$ Gyr. The star is chromospherically inactive with a log\,$R'_{HK}=-5.11\pm0.08$, leading to a rotational period of $26\pm3$ days. HD\,196067's visual companion is HD\,196068 (HIP\,102128), a G1 dwarf with an apparent magnitude V=7.18. The apparent separation of the pair is 16.8 arcsecond, which explains the two entries in the Hipparcos catalog. However, the two stars have very similar proper motions and are known to be gravitationally bound \citep[see e.g.][]{gould2004}. This conclusion is strengthened by the fact that the CORALIE average radial-velocities of the two stars are identical within uncertainties. The Hipparcos positions lead to a projected binary separation of 740 AU. This translates into an estimated binary semi-major axis of 932 AU. The spectroscopic analysis of HD\,196068 results in $T_{eff}=5997\pm44$, $[Fe/H]=0.34\pm0.04$, log\,$g=4.38\pm0.05$, and v$sin\,(i)=1.21$. These parameters lead to a stellar mass of M$_{*}=1.18\pm0.07$ M$_{\odot}$ with a poorly constrained age of $2.4\pm1.6$ Gyr, and a radius of $R= 1.34\pm0.23\,R_{\odot}$.

\subsection{HD\,219077 (HIP\,114699)}
HD\,219077 is a bright G8 dwarf with a magnitude $V= 6.12$ and an astrometric parallax of $\pi = 34.07\pm0.37$, which corresponds to an absolute magnitude of $M_{v}=3.78$ and a luminosity of 2.66 L$_{\odot}$. The star is chromospherically quiet with a mean stellar activity index of log\,$R'_{HK}=-5.12\pm0.01$, leading to a calibrated rotational period of $49\pm4$ days. The spectral analysis made by \cite{sousa2008} results in an effective temperature of $T_{eff}=5362\pm18$ and a stellar metallicity of $[Fe/H]=-0.13\pm0.01$, in agreement with the values published by \cite{ghezzi2010_feh}. Using the theoretical isochrones, we derived a mass of M$_{*}=1.05\pm0.02$ M$_{\odot}$ and an age of $8.9\pm0.3$ Gyr.

\subsection{HD\,220689 (HIP\,115662)}
HD\,220689 has a spectral type G3\,V with an astrometric parallax of  $\pi = 22.4\pm0.7$\,mas, which sets the star at a distance of $44.6\pm1.4$ pc from the Sun. The effective temperature and metallicity are $T_{eff}=5921\pm26$ and $[Fe/H]=0.00\pm0.03$. For this star we derived a mass M$_{*}=1.04\pm0.03$ M$_{\odot}$ with a luminosity of 1.24 L$_{\odot}$, an age of $3.5\pm1.9$ Gyr, and a radius of 1.07$\pm$0.04. The log\,$R'_{HK}$ of $-4.98\pm0.05$ derived from CORALIE spectra leads to a calibrated stellar rotation period of  $P_{rot}=20\pm3$ days.

\section{Radial velocities and orbital solutions}
\label{velocities}
The CORALIE echelle spectrograph was mounted on the 1.2\,m Euler Swiss telescope in June 1998. It went through a major hardware upgrade in June 2007 to increase overall efficiency. The reader is referred to \cite{segransan2009coralie} for more details. This hardware modification has affected the instrumental radial velocity zero point with offsets that depend on the target spectral type. For this reason, we decided to consider the upgraded spectrograph as a new instrument called CORALIE-07 (C07). We refer to the original CORALIE as CORALIE-98 (C98), and we adjust an offset between radial-velocities gathered before and after June 2007. 
Time series velocity data are fit with a Keplerian model using a genetic algorithm tool developed at the University of Geneva Observatory and called Yorbit \citep{segransan2011}. The interest of the genetic approach is to explore the different parameter spaces by creating a population of candidates to the Keplerian model and working toward better solutions. This tool has been developed and is especially useful for multiplanet systems, but it delivers accurate results in single-planet systems. This technique avoids the fitting algorithm becoming trapped in a local $\chi^{2}$ minimum. The best individuals of the evolved population are then used as input parameter to a Levenberg-Marquardt algorithm to obtain the best $\chi^{2}$ minimization fit between the assumed Keplerian model and the observed data. The parameter confidence intervals computed for a 68.3\% confidence level after 10\,000 Monte Carlo trials define the uncertainties. A Markov chain Monte Carlo (MCMC) algorithm is used to estimate the uncertainties of the fitted parameters when the orbital period is longer than the time span of the observations.\\
Because some of the planetary companions presented in this paper are fairly massive, the knowledge of the radial-velocity parameters can be used to search for a Hipparcos astrometric signature in the same way as in \cite{sahlmann2011}.  Except for HD\,10647b (see Sect.\,\ref{HD10647_section}) Hipparcos astrometry was not able to constrain the upper mass of the planets. For the lightest planetary companions, the astrometric signal is probably too weak, while for the longest orbital period, the comparatively short time span of the Hipparcos data implies poor orbit coverage (less than 25\%).
The standard orbital parameters derived from the best fitted Keplerian solution to the data are listed in Table\,\ref{table_orbits}. 

%
\begin{table*}
\caption{Single-planet Keplerian orbital solutions for HD\,27631, HD\,98649, HD\,106515A, HD\,166724, HD\,196067, and HD\,220689}
\label{table_orbits}
\tabcolsep=4.5pt      
\centering      
\begin{tabular}{l l c c c c c c c c}     
\hline\hline       
Parameters		&	 		&HD27631		&HD98649		&HD106515A		&HD166724		&HD196067		&HD219077		&HD220689		\\ 
\hline
$P$			&[days]			&2208$\pm$66		&4951$_{-465}^{+607}$	&3630$\pm$12		&5144$_{-467}^{+705}$	&3638$_{-185}^{+232}$	&5501$_{-119}^{+130}$	&2209$_{-81}^{+103}$	\\
$K$			&[m$^{-1} $]		&23.7$\pm$1.9		&176$_{-15}^{+44}$	&158.2$\pm$2.6		&71.0$\pm$1.7		&104$_{-24}^{+152}$	&181.4$\pm$0.8		&16.4$\pm$1.5		\\
$e$			&			&0.12$\pm$0.06		&0.85$\pm$0.05		&0.572$\pm$0.011	&0.734$\pm$0.020	&0.66$_{-0.09}^{+0.18}$	&0.770$\pm$0.003	&0.16$_{-0.07}^{+0.10}$	\\
$\omega$		&[deg]			&134$\pm$44		&248$\pm$9		&123.8$\pm$2.5		&202.3$\pm$3.6		&148.2$\pm$7.8		&57.61$\pm$0.42		&137$\pm$75		\\
$T_{0}$			&[JD]			&2453867$\pm$224	&2450271$\pm$22		&2451803.8$\pm$8.3	&2453212$\pm$556	&2453375$\pm$225	&2450455.8$\pm$1.4	&2455649$\pm$413	\\
\hline
$a_{1}\,$sin$\,i$	&[$10^{-3}$ AU]		&4.91$\pm$0.48		&36.9$\pm$3.4		&43.33$\pm$0.66		&22.1$\pm$2.0		&24.7$\pm$10.7		&58.5$\pm$1.3		&3.38$\pm$0.29		\\	
$f_{1}(m)$		&[$10^{-9}$ M$_{\odot}$]&3.3$\pm$0.9		&266$\pm$35		&824$\pm$37		&58.7$\pm$5.2		&152$\pm$87		&883$\pm$22		&1.09$\pm$0.25		\\
$m_{p}\,$sin$\,i$	&[M$_{\mathrm{Jup}}$]	&1.45$\pm$0.14		&6.8$\pm$0.5		&9.61$\pm$0.14		&3.53$\pm$0.11		&6.9$_{-1.1}^{+3.9}$	&10.39$\pm$0.09		&1.06$\pm$0.09		\\
$a$			&[AU]			&3.25$\pm$0.07		&5.6$\pm$0.4		&4.590$\pm$0.010	&5.42$\pm$0.43		&5.02$\pm$0.19		&6.22$\pm$0.09		&3.36$\pm$0.09		\\
\hline
$\gamma_{C98}$		&[kms$^{-1} $] 		&21.1133		&4.2671			&20.8095	   	&-17.6187		&-10.8887		&-30.8691		&12.1928		\\
			&			&$\pm$0.0049		&$\pm$0.0075		&$\pm$0.0026		&$\pm$0.0029		&$\pm$0.0090		&$\pm$0.0033		&$\pm$0.0036		\\
$\Delta V_{C07-C98}$	&[ms$^{-1} $] 		&-2.5$\pm$4.9	 	&13.5$\pm$8.4		&-5.8$\pm$4.7	 	&-2.8$\pm$3.2		&2.6$\pm$4.1		&7.5$\pm$2.6		&0.0$\pm$3.5		\\
$\Delta V_{HARPS-C98}$	&[ms$^{-1} $] 		&			&			&			&22.6$\pm$2.8		&			&41.7$\pm$2.2		&			\\
\hline
$\mathrm{N_{mes}}$ 	&			&60 (23/37)		&43 (11/32)		&46 (22/24)		&102 (18/29/55)		&82 (30/52)		&93 (36/27/30)		&44 (11/33)		\\
$\Delta T$		&[years]		&12.8			&9.4			&13.2			&11.0			&13.4			&13.3			&12.1			\\
$\chi_{r}^{2}$		&			&1.37$\pm$0.23		&1.35$\pm$0.27		&1.55$\pm$0.28		&1.24$\pm$0.16		&2.48$\pm$0.26		&1.15$\pm$0.16		&1.14$\pm$0.25		\\  
G.o.F			&			&1.77			&1.42			&2.16			&1.56			&6.55			&0.98			&0.66			\\
$\sigma_{(O-C)}$ 	&[ms$^{-1}$]		&7.28			&6.64			&7.56			&3.76			&9.72			&1.88			&6.08			\\     
$\sigma_{(O-C)_{C98}}$ 	&[ms$^{-1}$]		&6.31			&10.31			&9.63			&10.61			&8.88			&6.18			&5.47			\\     
$\sigma_{(O-C)_{C07}}$ 	&[ms$^{-1}$]		&7.56			&5.68			&6.66			&8.30			&9.92			&7.63			&6.18			\\    
$\sigma_{(O-C)^{HARPS}}$&[ms$^{-1}$]		&			&			&			&3.57			&			&1.53			&			\\			
\hline                  
\end{tabular}
\tablefoot{
Confidence intervals are computed for a 68.3\% confidence level after 10\,000 Monte Carlo trials. $C98$ stands for CORALIE-98 and $C07$ for CORALIE-07. $\Delta T$ is the time interval between the first and last measurements, $\chi_{r}^{2}$ is the reduced $\chi^{2}$, G.o.F. is the goodness of fit and $\sigma_{(O-C)}$ the weighted rms of the residuals around the derived solution.
}
\end{table*}

   \begin{figure}
   \centering
   \includegraphics[width=9.2cm]{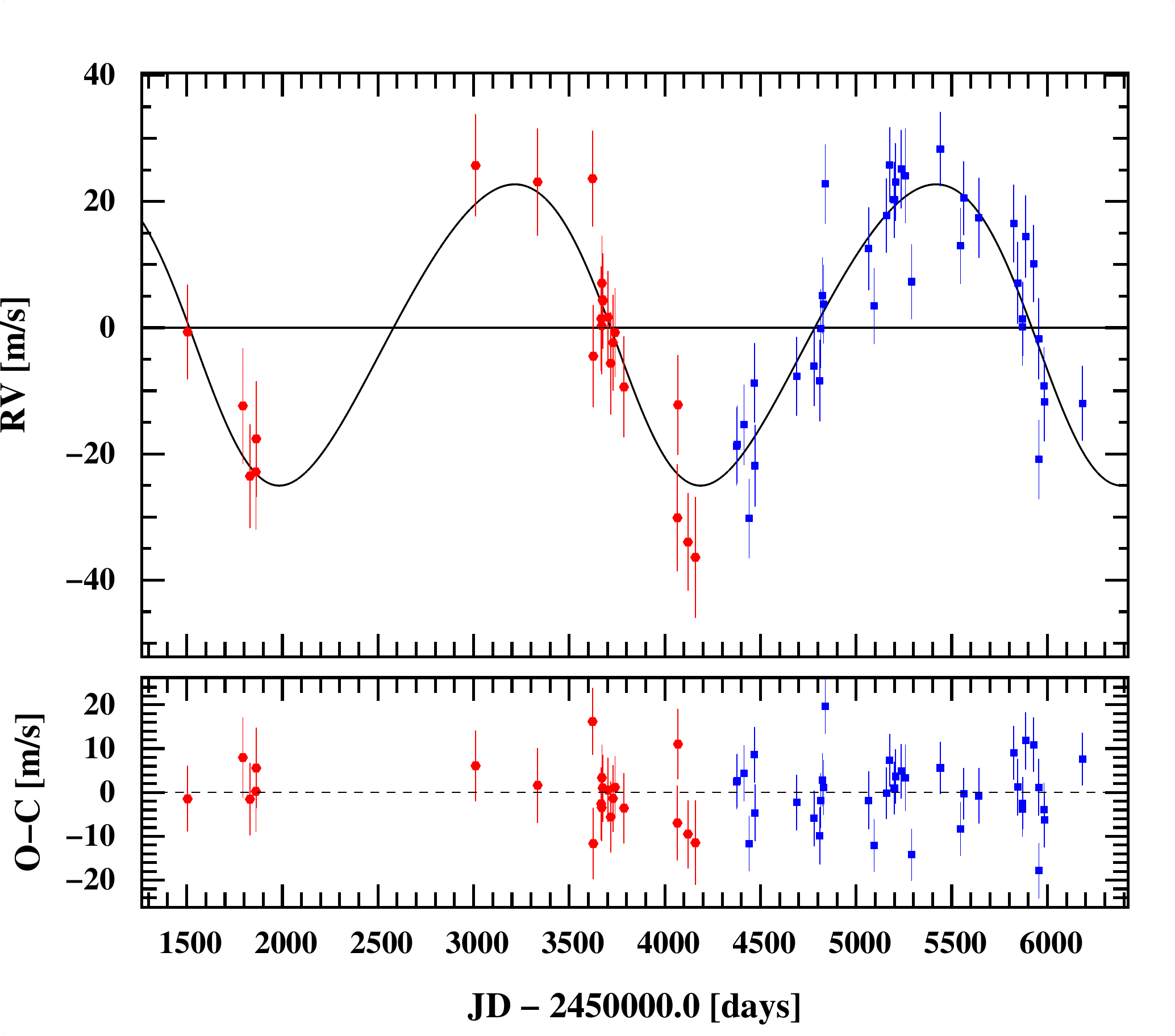}
   \includegraphics[width=9.2cm]{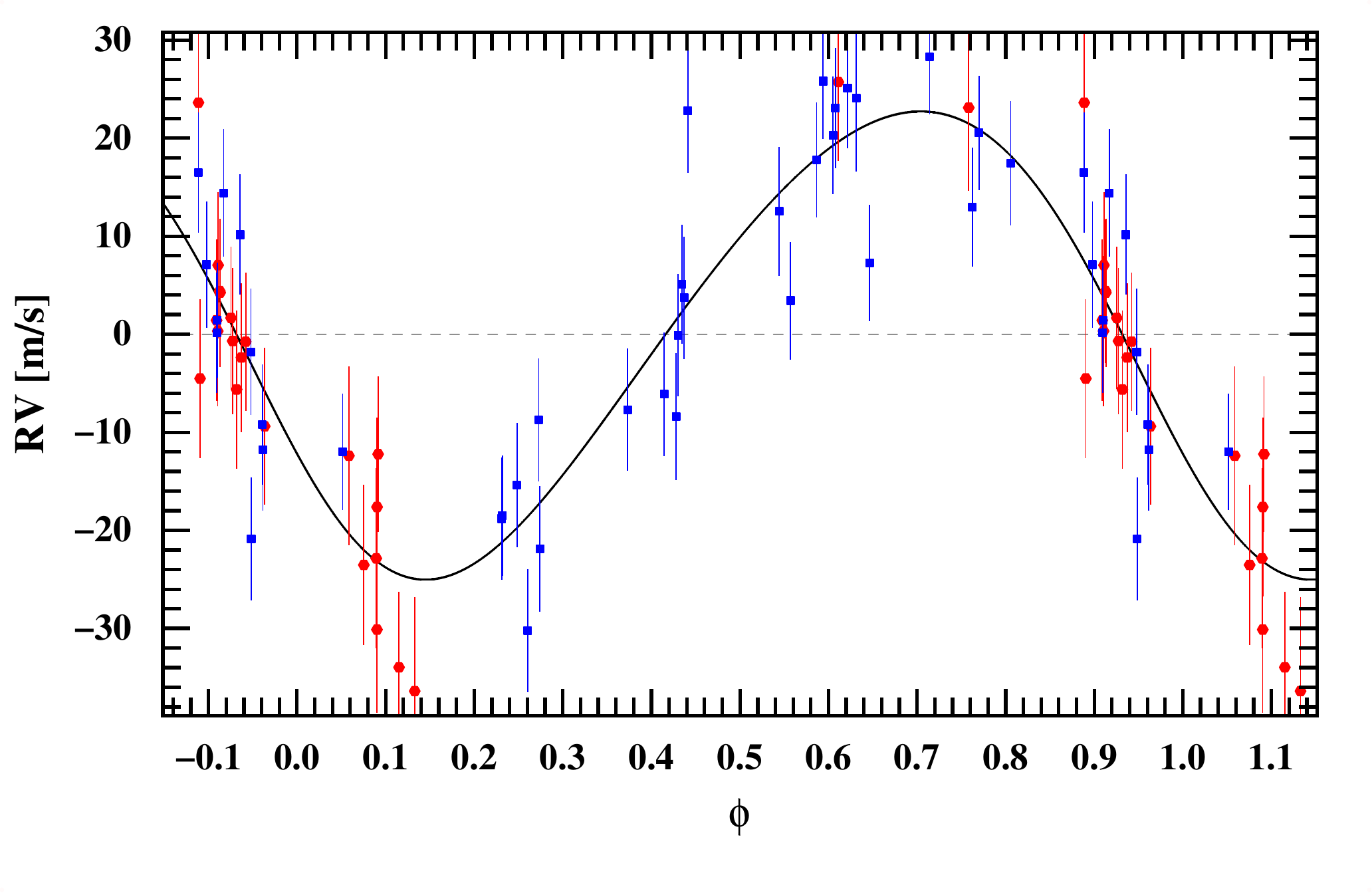}
      \caption{Radial-velocity measurements as a function of Julian Date obtained with CORALIE-98 (red)
      and CORALIE-07 (blue) for HD\,27631. The best single-planet 
      Keplerian model is represented as a black curve. The residuals are displayed at the bottom
      of the top figure and the phase-folded radial-velocity measurements 
      are displayed in the bottom diagram.}
       \label{HD27631_orb}
   \end{figure}
   \begin{figure}
   \centering
   \includegraphics[width=9.2cm]{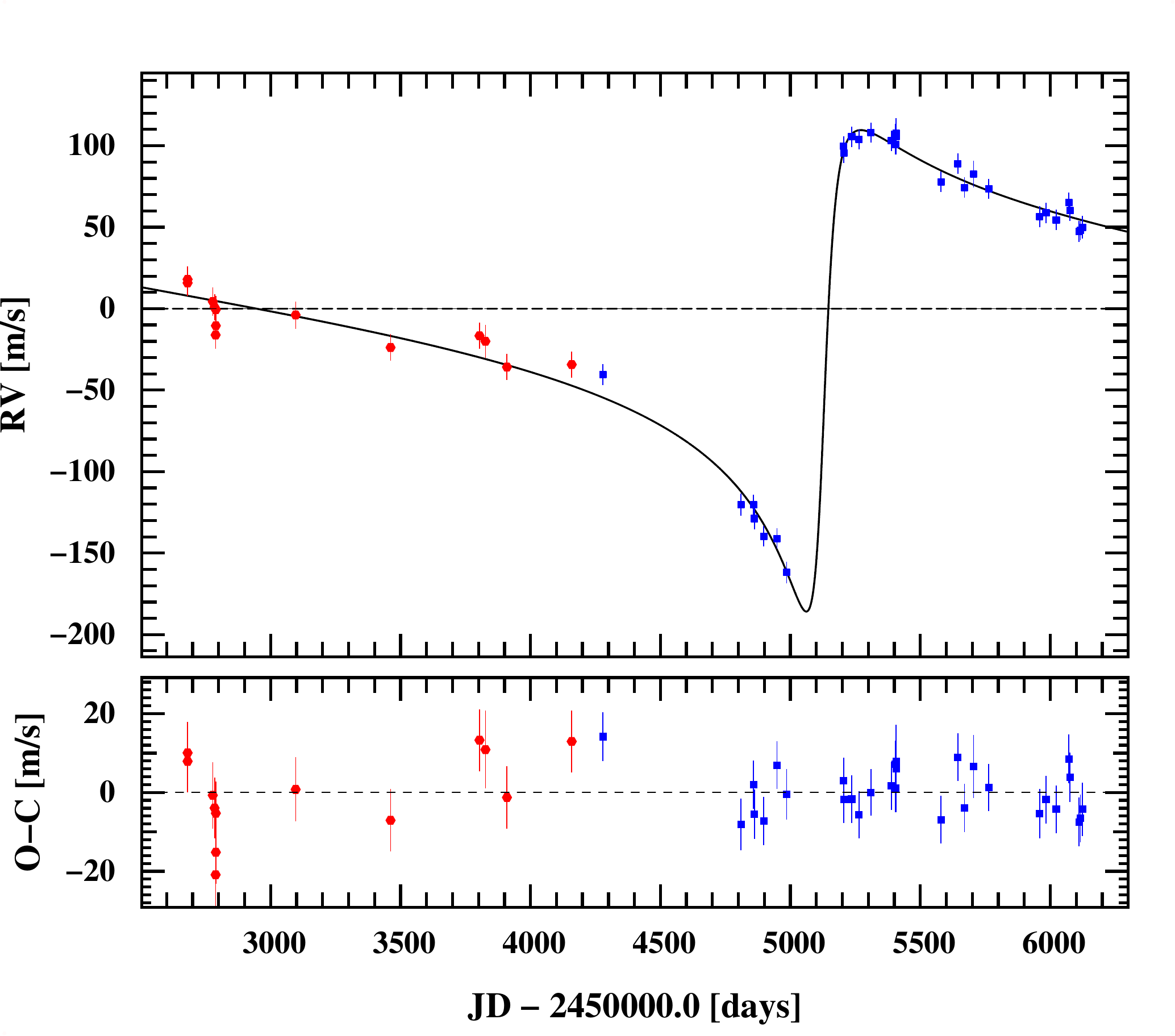}
   \includegraphics[width=9.2cm]{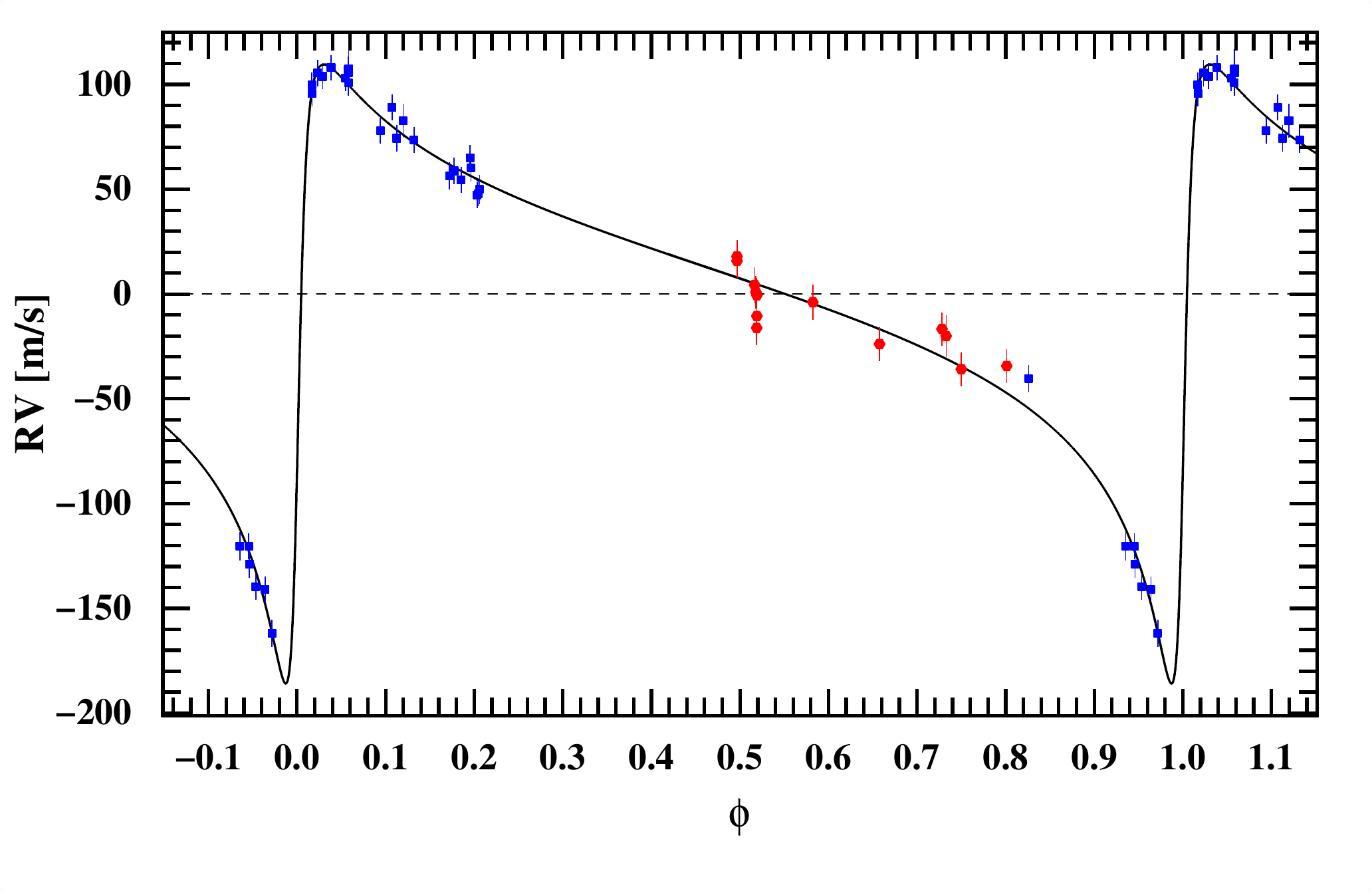}
      \caption{Radial-velocity measurements as a function of Julian Date obtained with CORALIE-98 (red)
      and CORALIE-07 (blue) for HD\,98649. The best single-planet 
      Keplerian model is represented as a black curve. The residuals are displayed at the bottom
      of the top figure and the phase-folded radial-velocity measurements 
      are displayed in the bottom diagram.}
       \label{HD98649_orb}
   \end{figure}
   \begin{figure}
   \centering
   \begin{minipage}[]{.4\linewidth}
    \begin{center}
       \includegraphics[width=5.3cm]{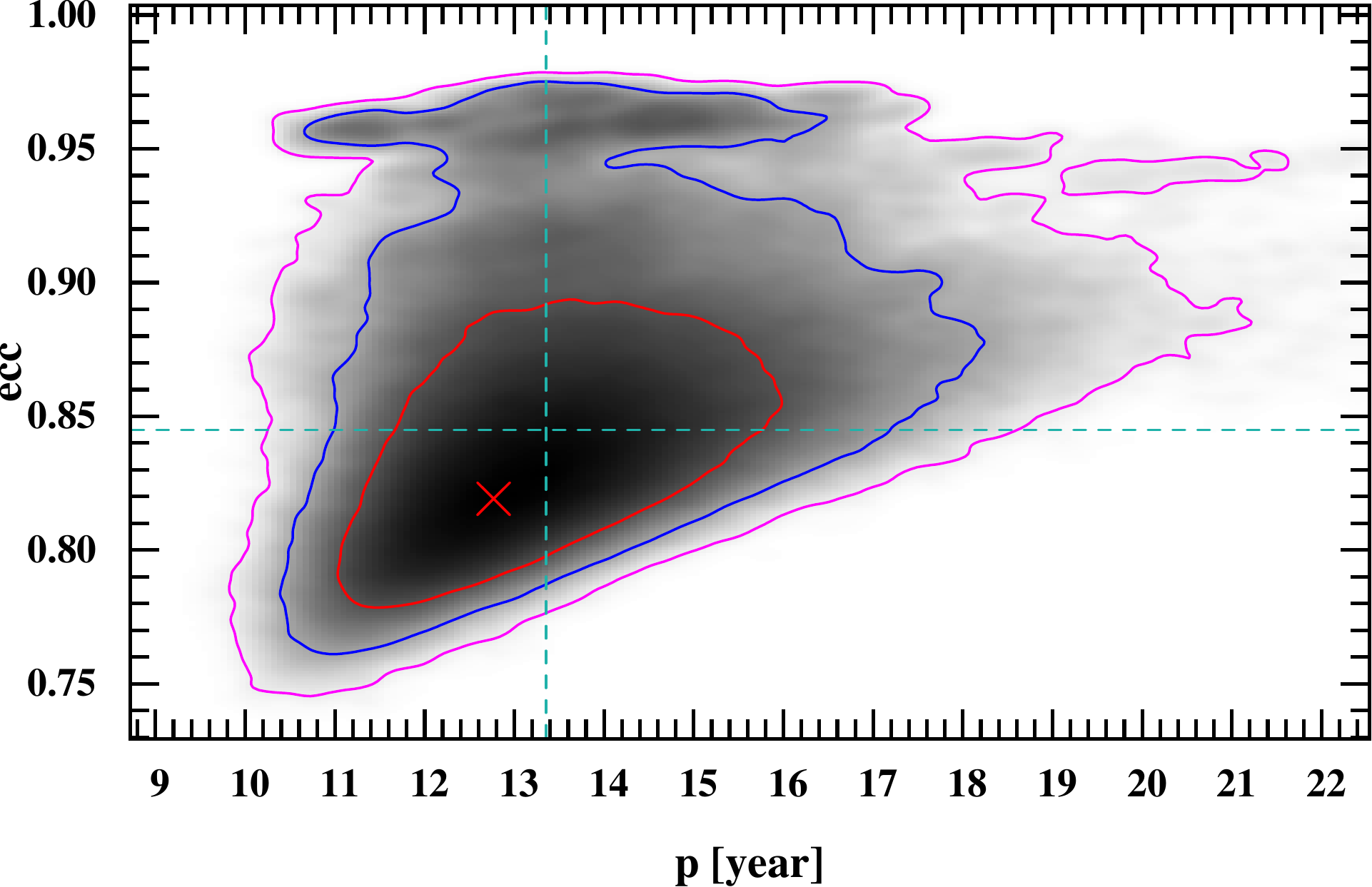}
    \end{center}
   \end{minipage}
   \hfill
   \begin{minipage}[]{.4\linewidth}
    \begin{center}
       \includegraphics[width=3.3cm]{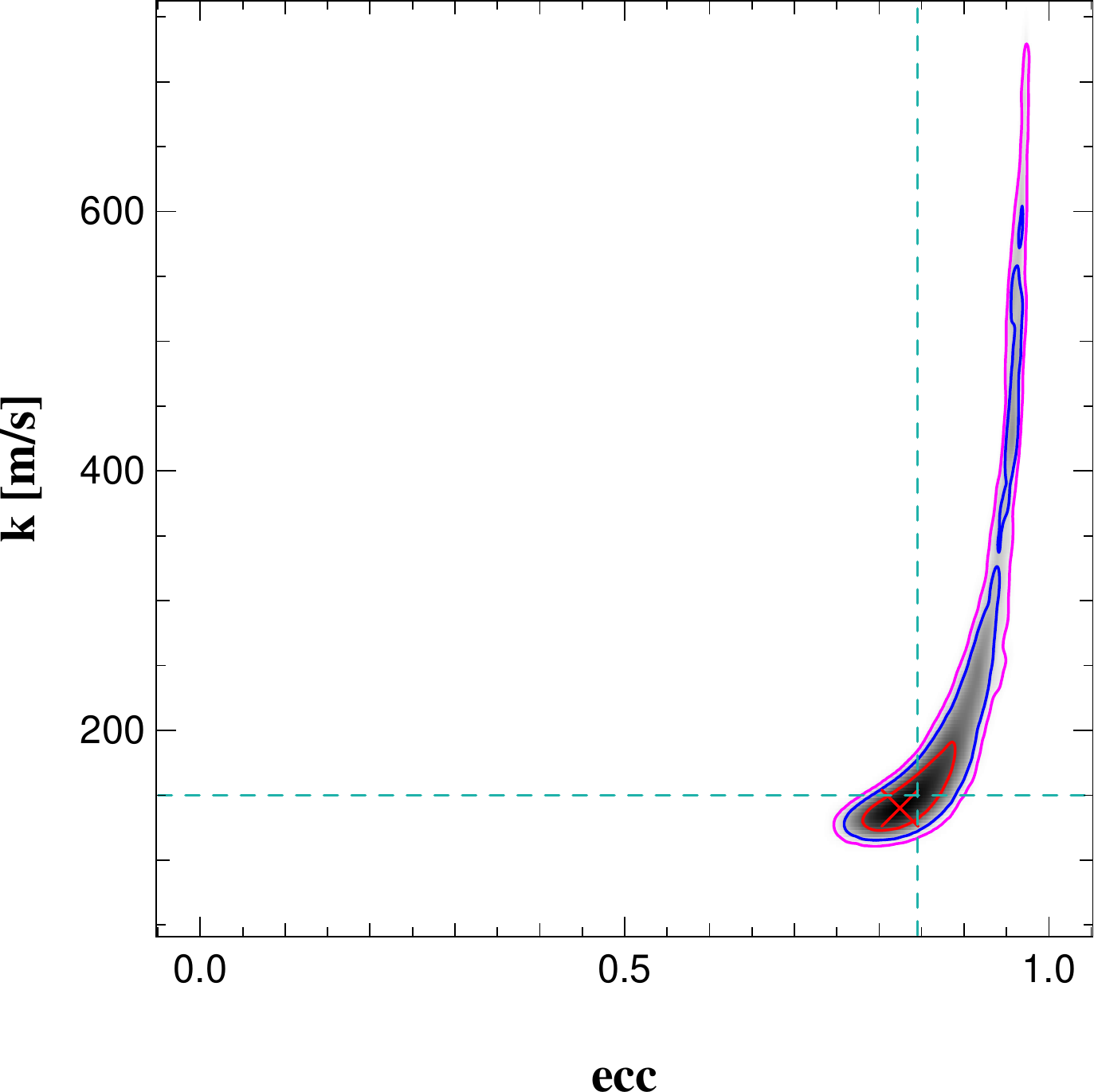}
    \end{center}
   \end{minipage}
   \begin{minipage}[]{.4\linewidth}
    \begin{center}
       \includegraphics[width=5.3cm]{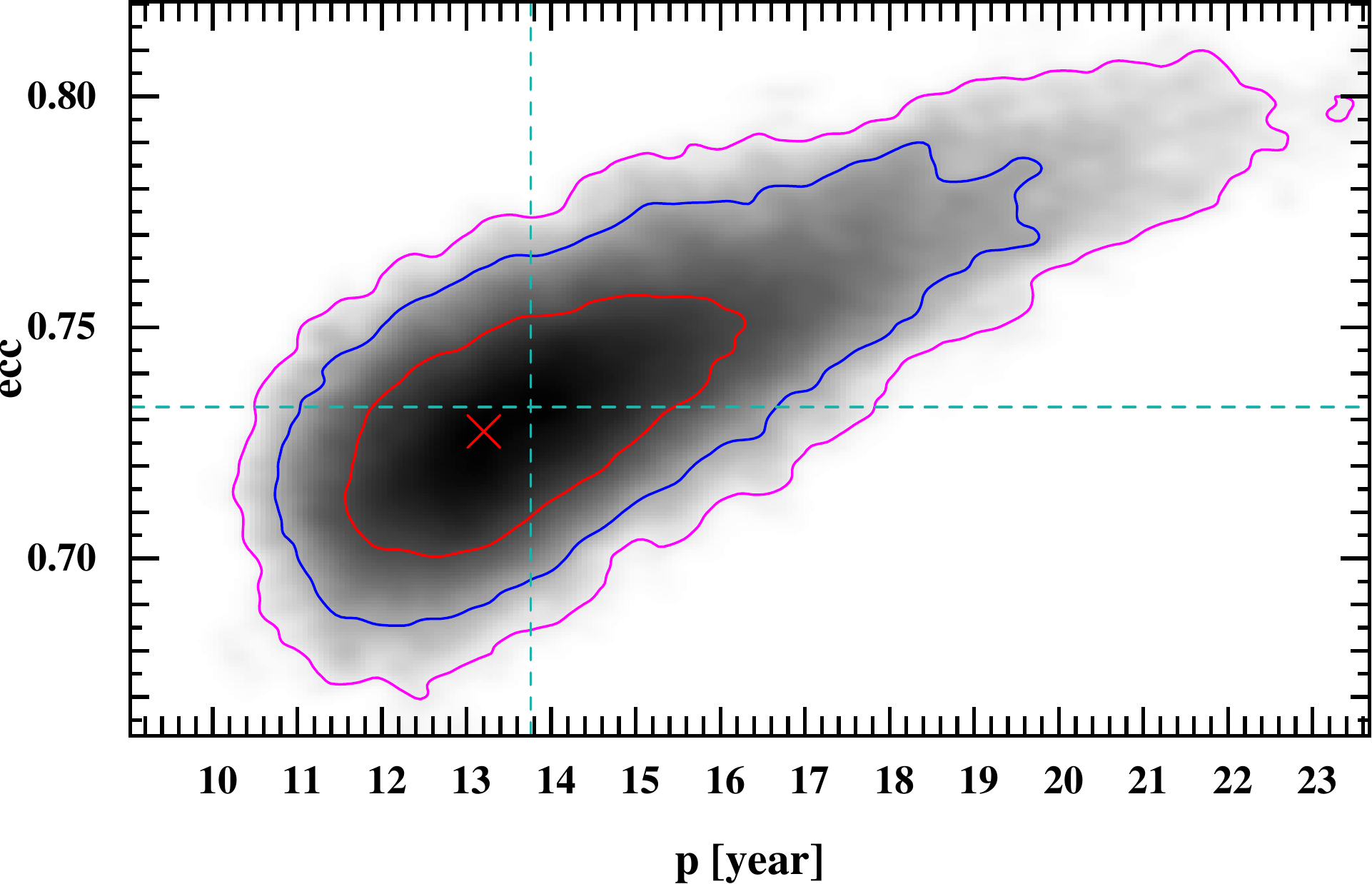}
    \end{center}
   \end{minipage}
   \hfill
   \begin{minipage}[]{.4\linewidth}
    \begin{center}
       \includegraphics[width=3.3cm]{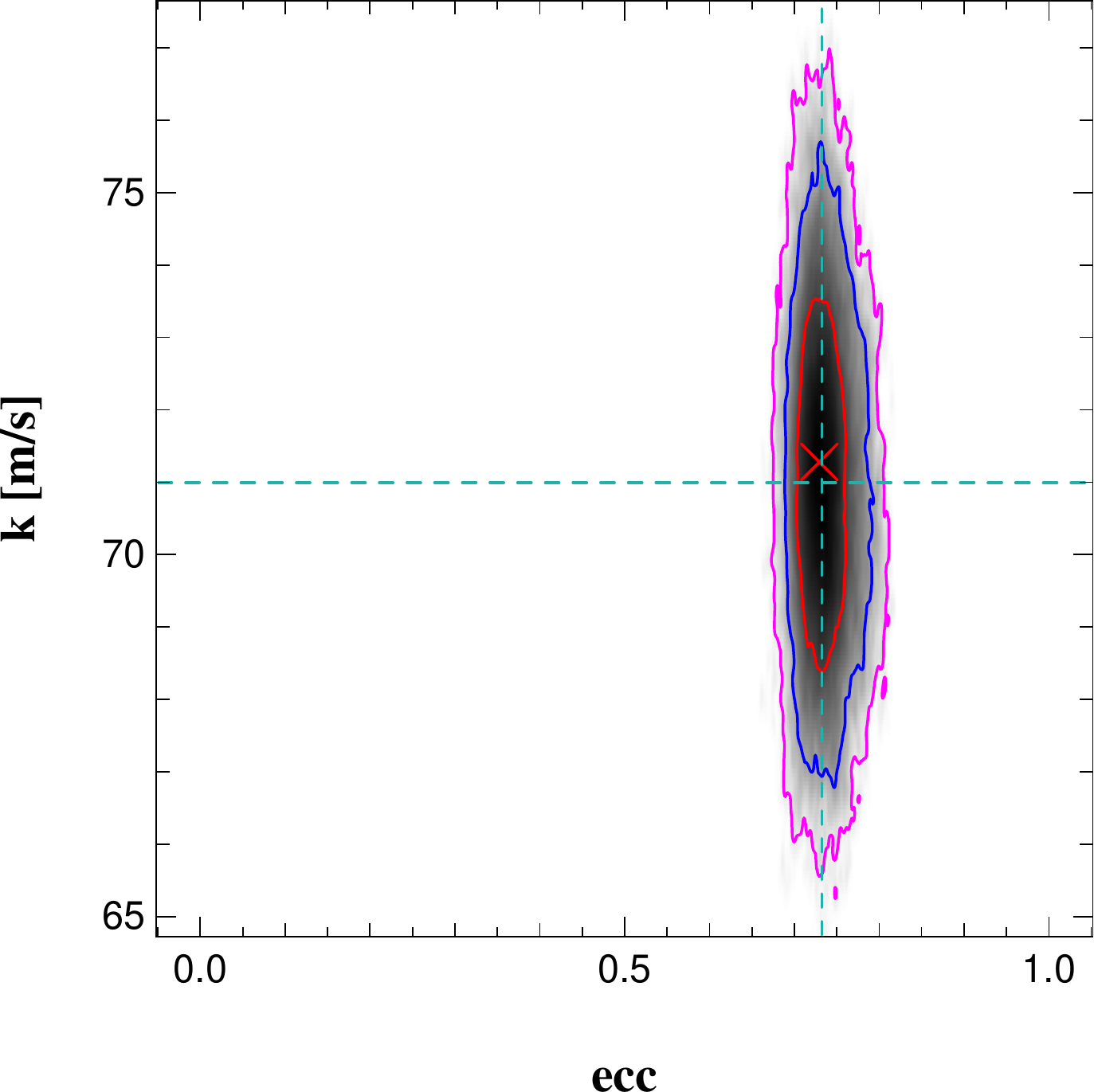}
    \end{center}
   \end{minipage}
   \begin{minipage}[]{.4\linewidth}
    \begin{center}
       \includegraphics[width=5.3cm]{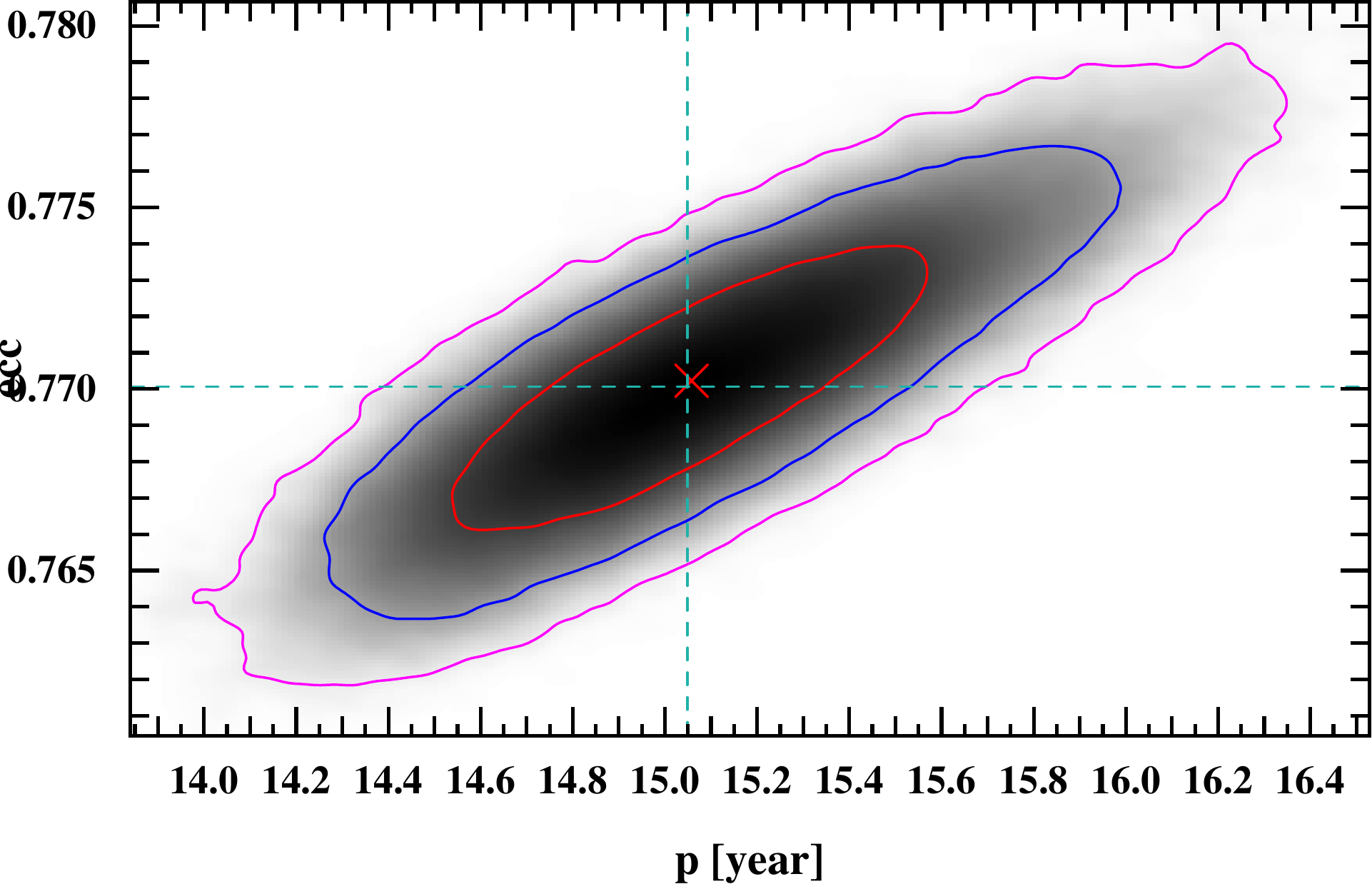}
    \end{center}
   \end{minipage}
   \hfill
   \begin{minipage}[]{.4\linewidth}
    \begin{center}
       \includegraphics[width=3.3cm]{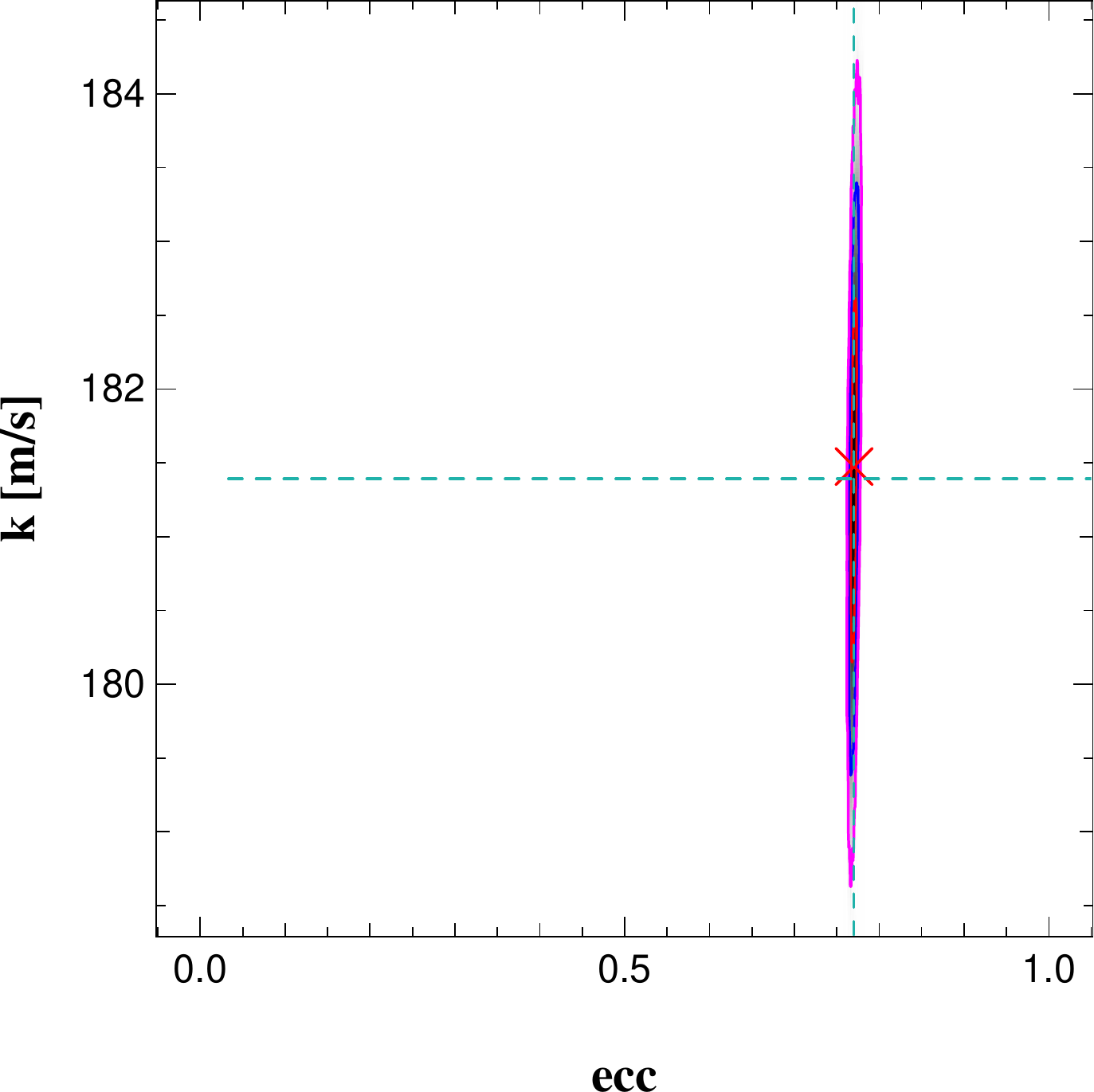}
    \end{center}
   \end{minipage}
   \caption{Covariance between orbital period, semi-amplitude, and eccentricity for HD\,98649b (top), HD\,166724b (middle), and HD\,219077b (down) as characterized after 5x10$^{6}$ Markov-chain Monte Carlo iterations}
   \label{correl_ecc_p_K1}
   \end{figure}
   \begin{figure}
   \centering
   \includegraphics[width=9.2cm]{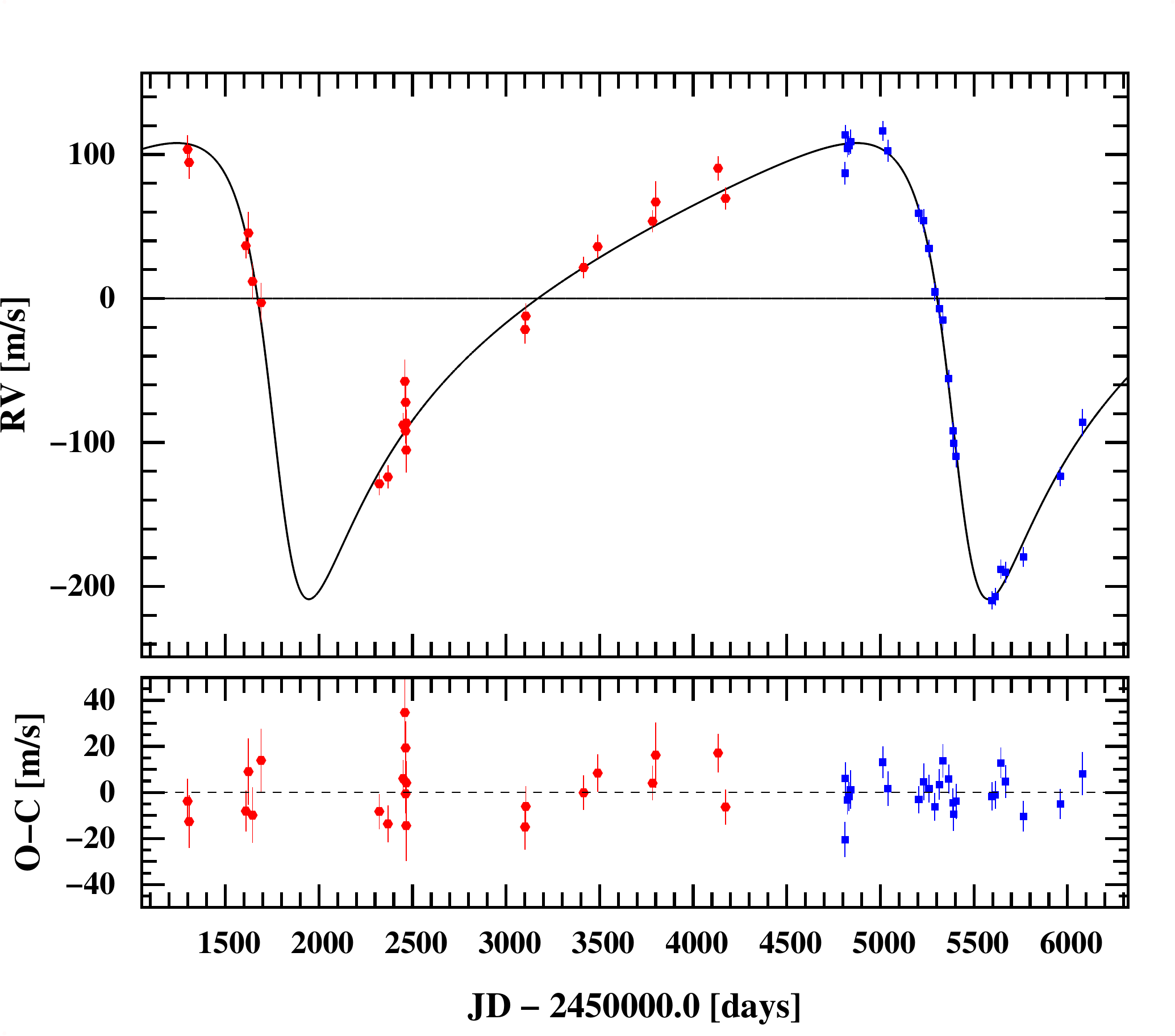}
   \includegraphics[width=9.2cm]{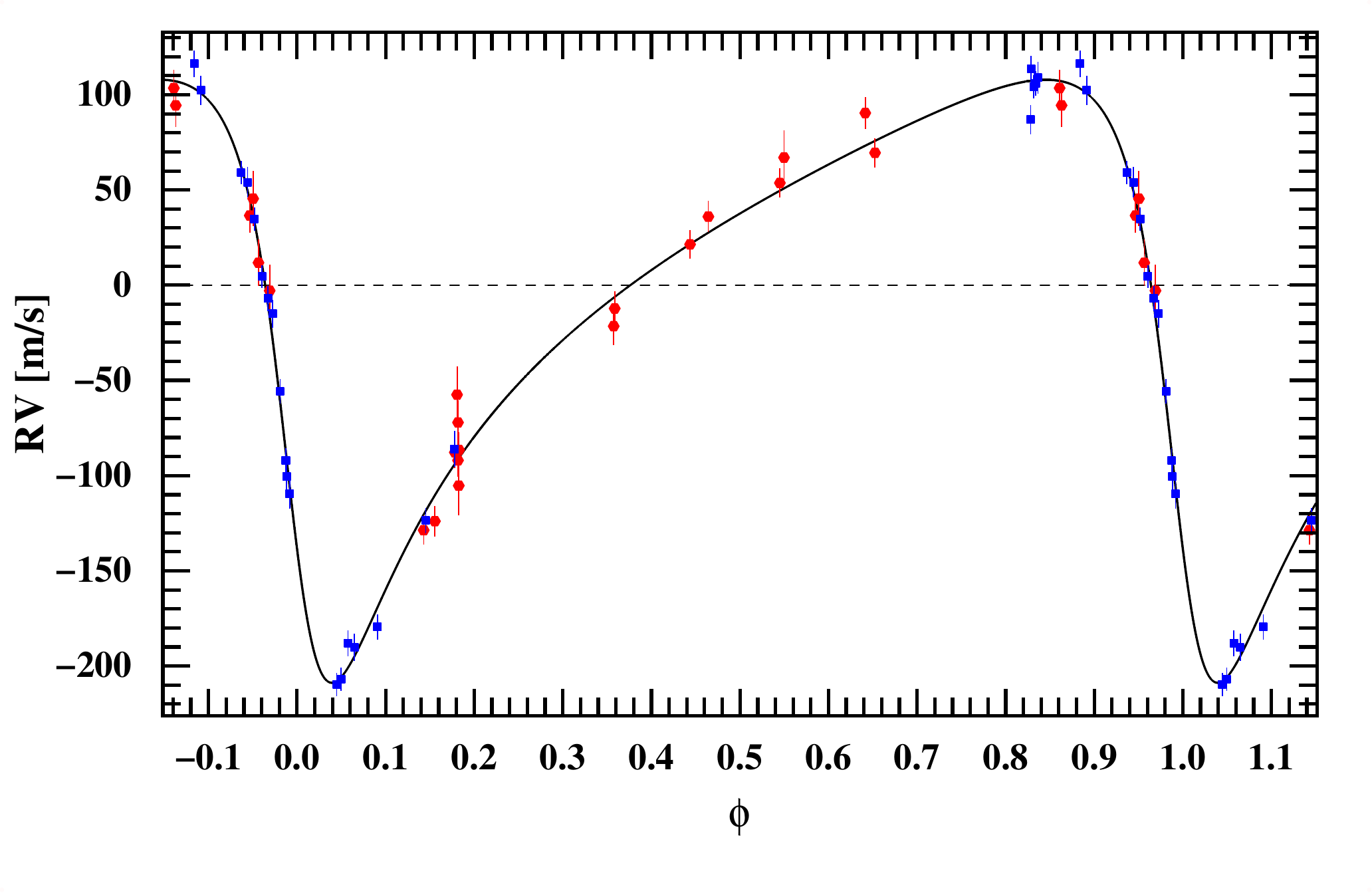}
      \caption{Radial-velocity measurements as a function of Julian Date obtained with CORALIE-98 (red)
      and CORALIE-07 (blue) for HD\,106515A. The best single-planet 
      Keplerian model is represented as a black curve. The residuals are displayed at the bottom
      of the top figure and the phase-folded radial-velocity measurements 
      are displayed in the bottom diagram.}
       \label{HD106515A_orb}
   \end{figure}
   \begin{figure}
   \centering
   \includegraphics[width=9.2cm]{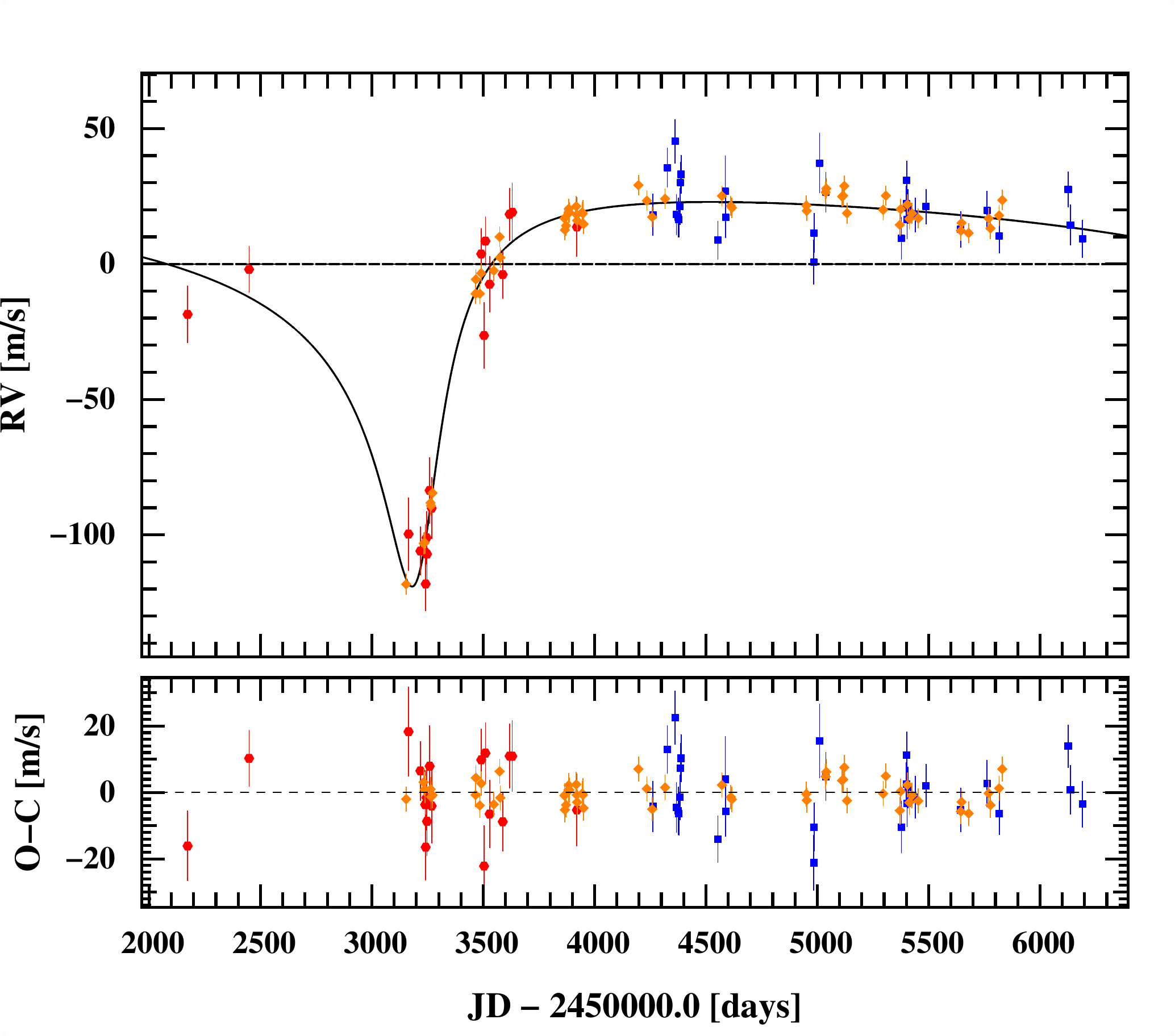}
   \includegraphics[width=9.2cm]{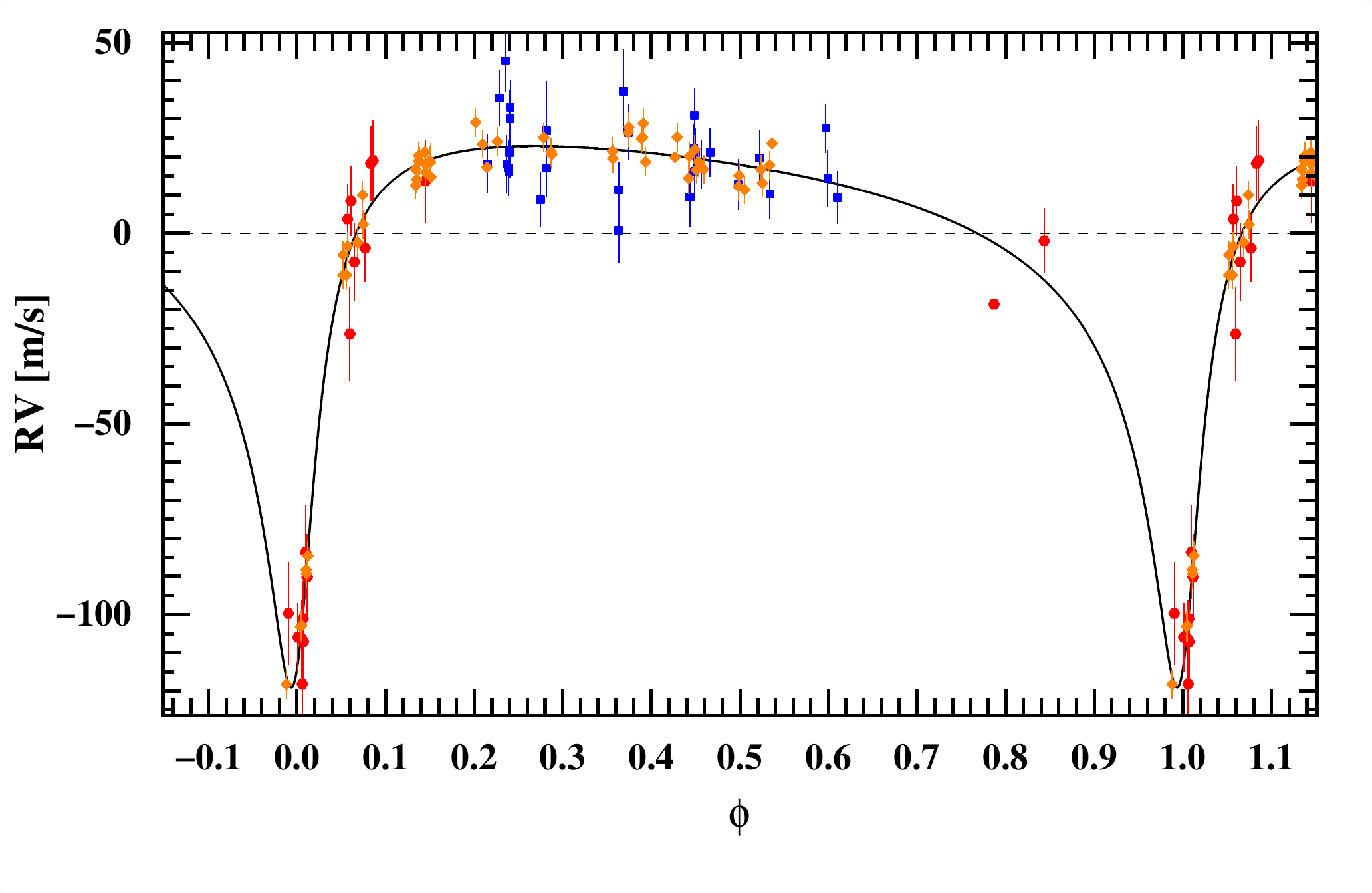}
      \caption{Radial-velocity measurements as a function of Julian Date obtained with CORALIE-98 (red),
      CORALIE-07 (blue), and HARPS (yellow) for HD\,166724. The best single-planet 
      Keplerian model is represented as a black curve. The residuals are displayed at the bottom
      of the top figure and the phase-folded radial-velocity measurements 
      are displayed in the bottom diagram.}
       \label{HD166724_orb}
   \end{figure}
   \begin{figure}
   \centering
   \includegraphics[width=9.2cm]{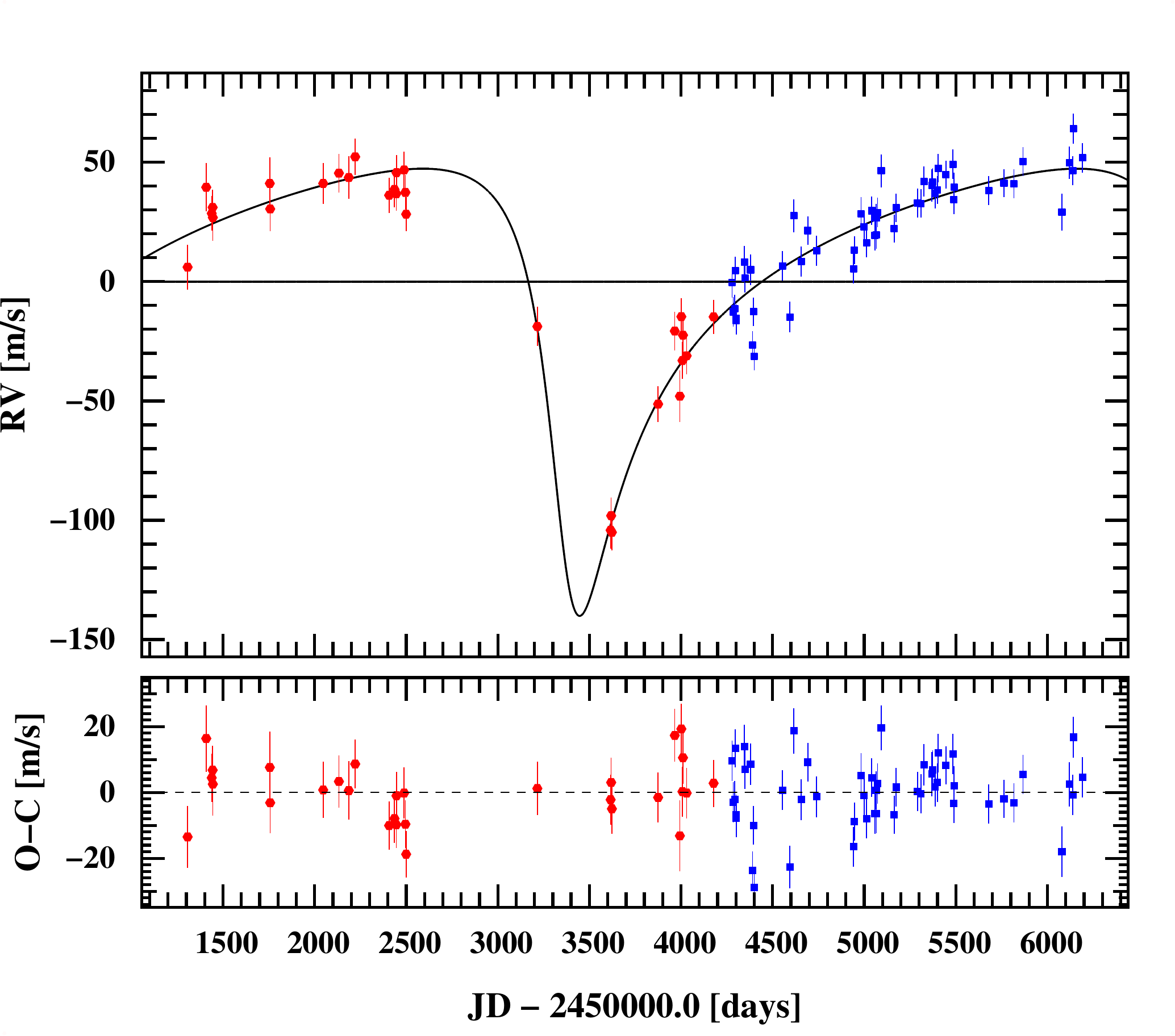}
   \includegraphics[width=9.2cm]{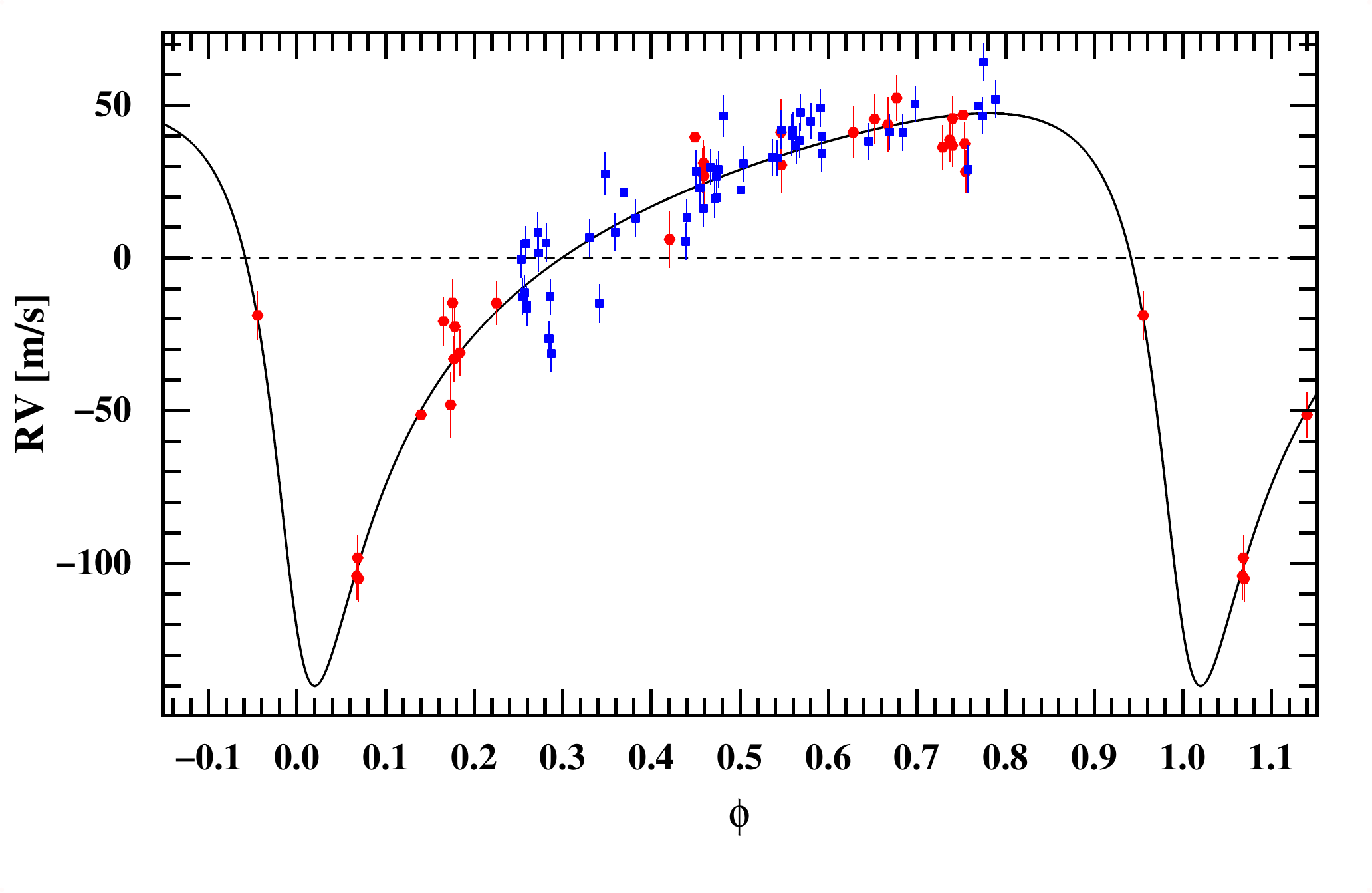}
      \caption{Radial-velocity measurements as a function of Julian Date obtained with CORALIE-98 (red)
      and CORALIE-07 (blue) for HD\,196067. The best single-planet 
      Keplerian model is represented as a black curve. The residuals are displayed at the bottom
      of the top figure and the phase-folded radial-velocity measurements 
      are displayed in the bottom diagram.}
       \label{HD196067_orb}
   \end{figure}
   \begin{figure}
   \centering
   \includegraphics[width=9.2cm]{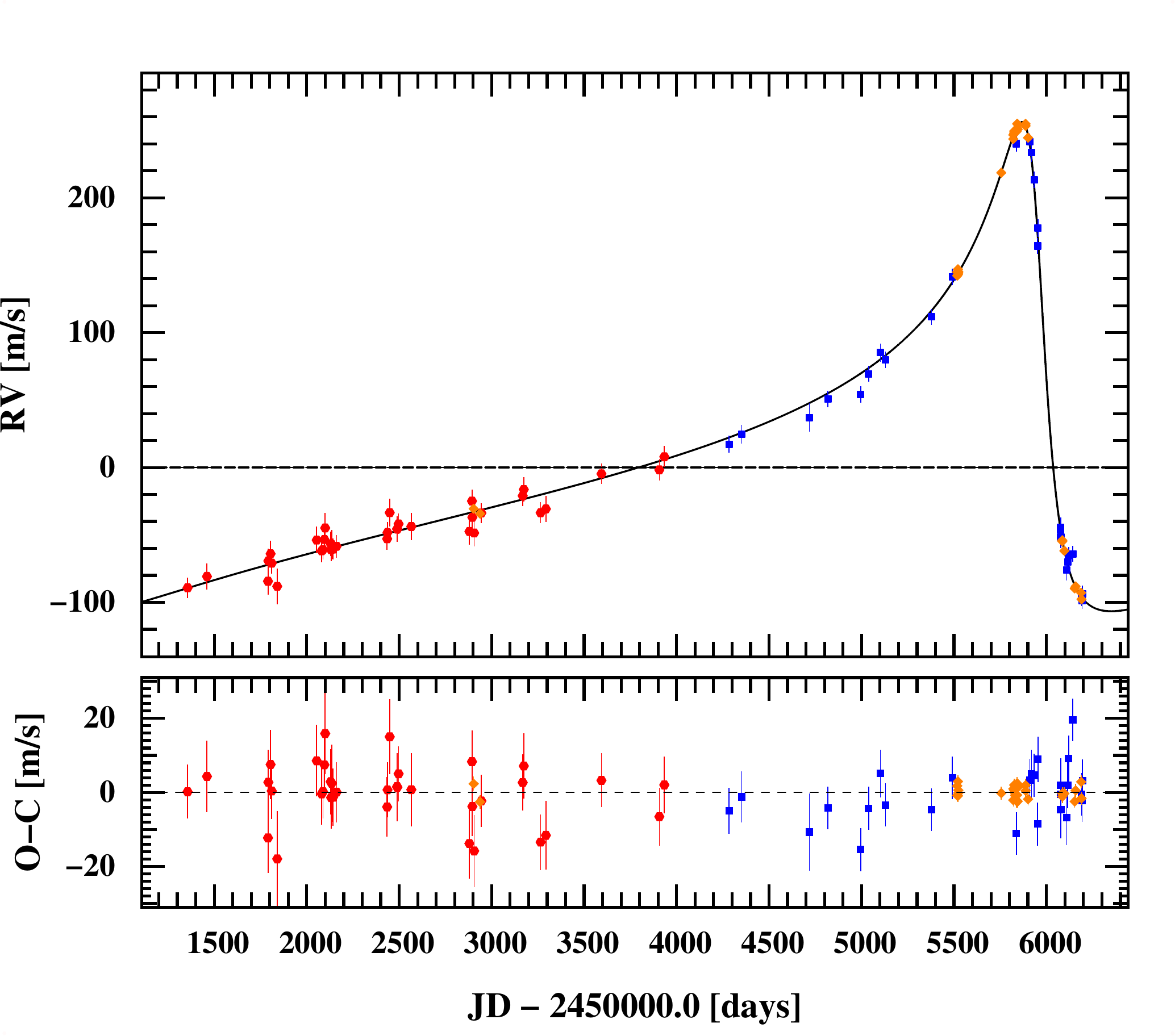}
   \includegraphics[width=9.2cm]{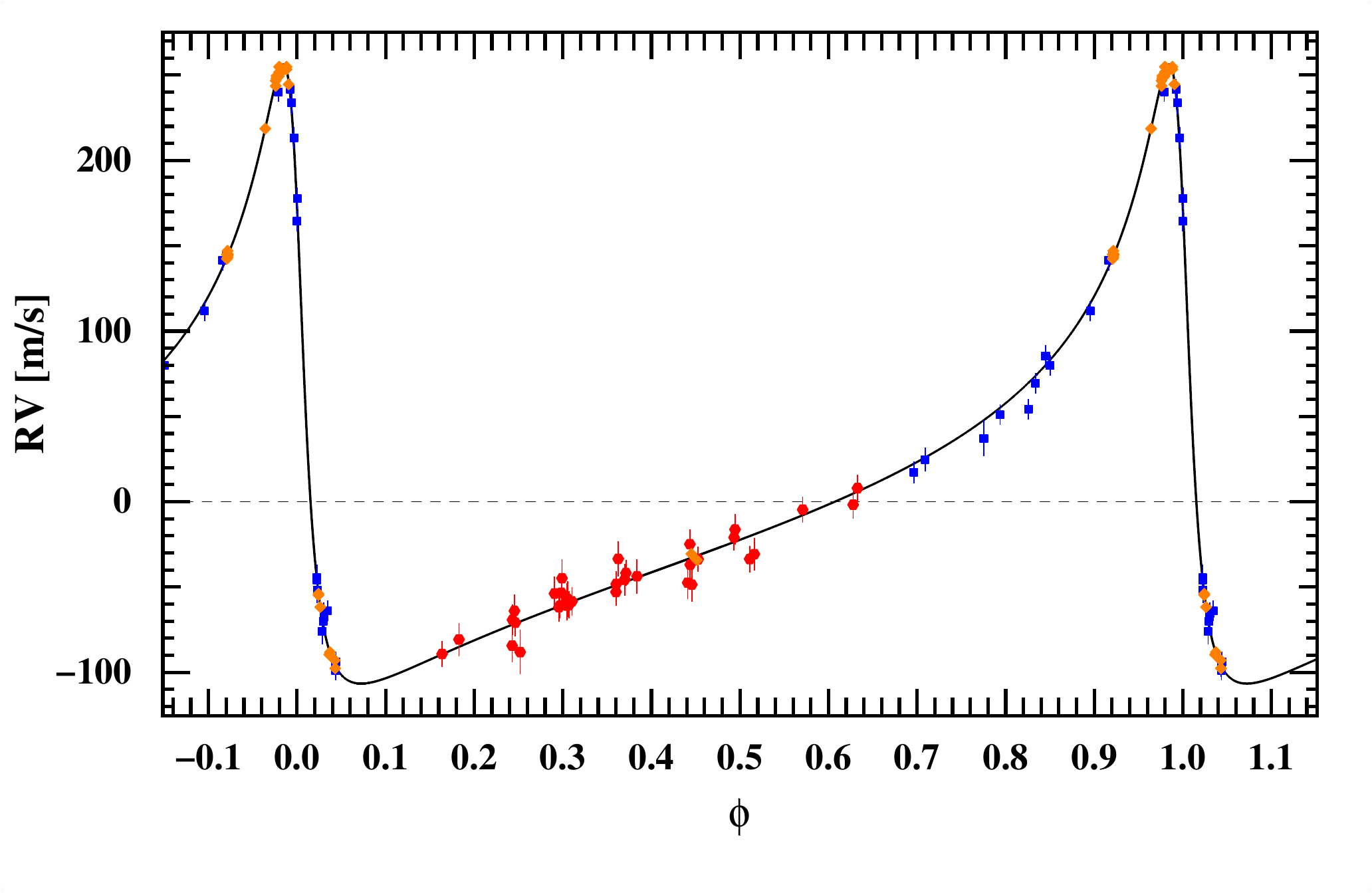}
      \caption{Radial-velocity measurements as a function of Julian Date obtained with CORALIE-98 (red), 
      CORALIE-07 (blue) and HARPS (yellow) for HD\,219077. The best single-planet 
      Keplerian model is represented as a black curve. The residuals are displayed at the bottom
      of the top figure and the phase-folded radial-velocity measurements 
      are displayed in the bottom diagram.}
       \label{HD219077_orb}
   \end{figure}
   \begin{figure}
   \centering
   \includegraphics[width=9.2cm]{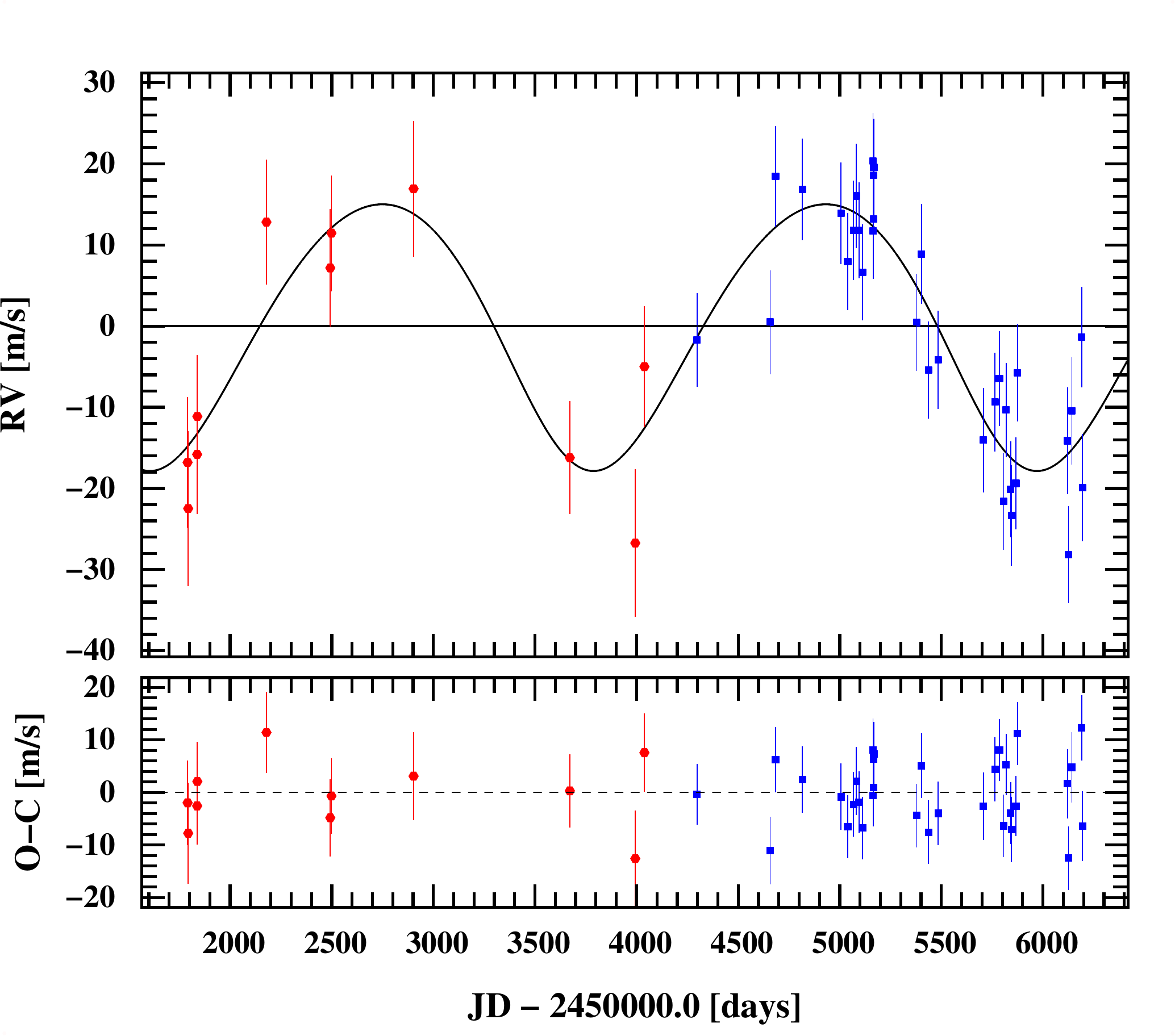}
   \includegraphics[width=9.2cm]{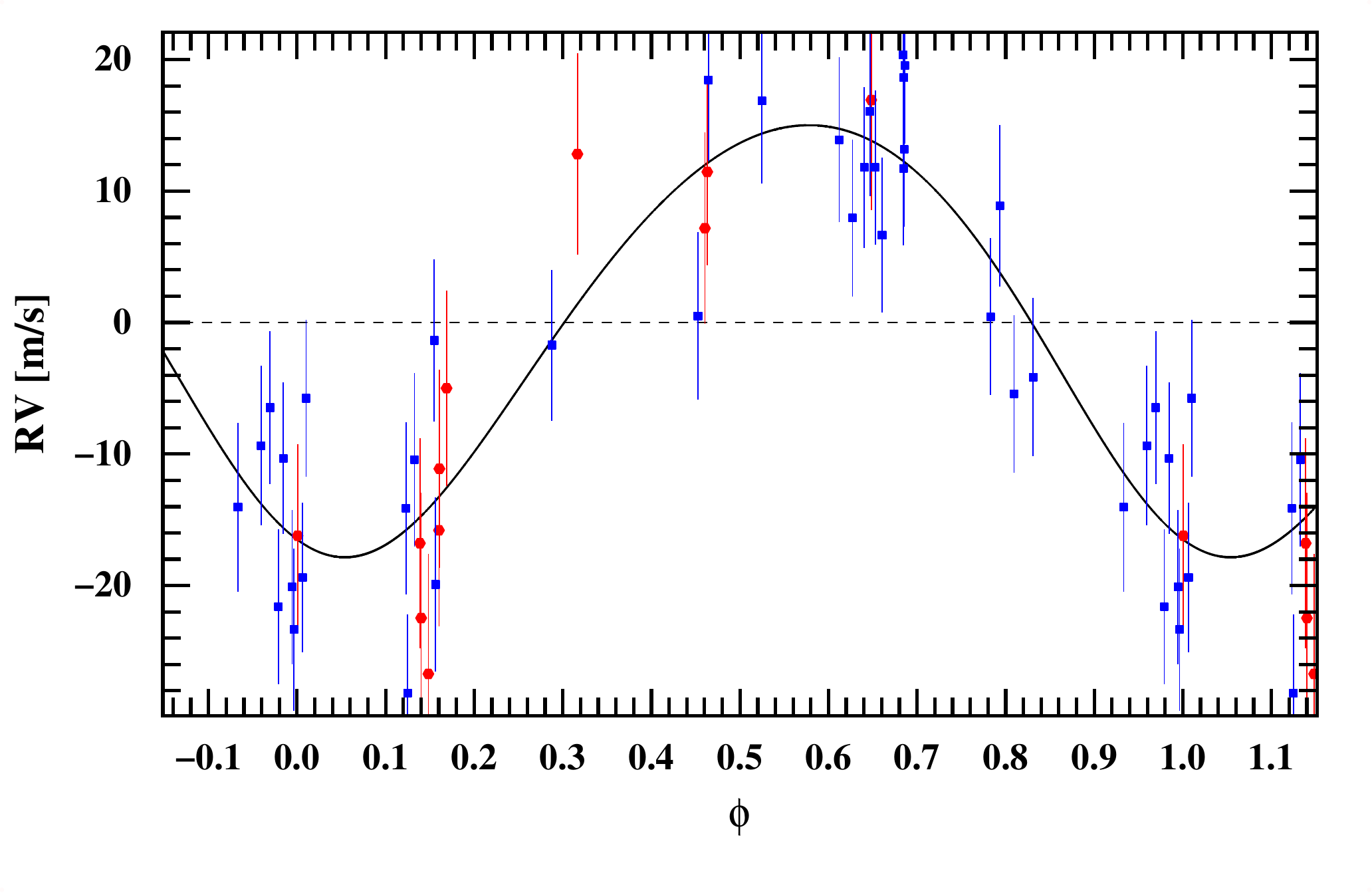}
      \caption{Radial-velocity measurements as a function of Julian Date obtained with CORALIE-98 (red)
      and CORALIE-07 (blue) for HD\,220689. The best single-planet 
      Keplerian model is represented as a black curve. The residuals are displayed at the bottom
      of the top figure and the phase-folded radial-velocity measurements 
      are displayed in the bottom diagram.}
       \label{HD220689_orb}
   \end{figure}

\subsection{A six-year period Jovian planet around HD\,27631}
We have acquired 60 radial velocities measurements of HD\,27631 since November 1999. Twenty-three of these were obtained with CORALIE-98 with a typical signal-to-noise ratio of 40 (per pixel at 550\,nm) leading to a mean measurement uncertainty including $\gamma$-noise and calibration errors of 7.8 ms$^{-1}$. Thirty-seven additional radial-velocity measurements were obtained with CORALIE-07 with a mean signal-to-noise ratio of 75 leading to a mean measurement uncertainty of 6.1\,ms$^{-1}$.
Figure\,\ref{HD27631_orb} shows the CORALIE radial velocities and the corresponding best-fit Keplerian model. Since each data set has its own velocity zero point, the velocity offset between CORALIE-98 and CORALIE-07 is an additional free parameter to include in the model, as described before. The resulting orbital parameters are $P$ = 2208$\pm$66 days, $e$ = 0.12$\pm$0.06, and $K$ = 23.7$\pm$1.9 m$^{-1}$, implying a minimum mass $m_{p}\,$sin$\,i$ = 1.45$\pm$0.14 M$_{\mathrm{Jup}}$ orbiting with a semi-major axis $a=3.25\pm0.07$ AU. The rms to the Keplerian fit is 6.31\,ms$^{-1}$ for CORALIE-98 and 7.56\,ms$^{-1}$ for CORALIE-07, yielding a reduced $\chi^{2}$ of 1.37$\pm$0.23 and 1.77 as goodness of fit. The single-planet Keplerian model thus adequately explains the radial-velocity variation, because for each instrument the rms is compatible with the mean measurement uncertainties.

\subsection{A 7 Jupiter-mass companion on an eccentric orbit of 14 years around HD\,98649}
Eleven radial-velocity measurements of HD\,98649 were obtained between February 2003 and March 2007 with CORALIE-98 showing a slight linear drift of -9.7 ms$^{-1}$yr$^{-1}$. The average measurement uncertainty is 7.7 ms$^{-1}$ for a mean signal-to-noise ratio of 48. Since the trend was very constant and the radial-velocity dispersion did not show any sign of a potential planetary companion, the observing frequency was decreased. In December 2008 after nearly six years of observation, a new measurement made with CORALIE-07 revealed a radial-velocity 70 ms$^{-1}$ below the expected value.\\
Additional radial-velocities gathered with CORALIE-07 unveiled a companion with a minimum mass $m_{p}\,$sin$\,i$ = 6.8$\pm$0.5 M$_{\mathrm{Jup}}$ on an eccentric orbit e = 0.85$\pm$0.05 around HD\,98649 with a period of $P$ = 4951$_{-465}^{+607}$ days, implying a semi-major axis $a=5.6\pm0.4$ AU. The thirty-two radial-velocities gathered with CORALIE-07 have a typical signal-to-noise ratio of 85, leading to a mean measurement uncertainty of 5.9\,ms$^{-1}$. With an observing time span of 9.4 years the radial velocity measurements cover only 70\% of orbital phase, explaining why the period is poorly constrained. However, the good data coverage near the periastron passage of the planet allows us to constrain the orbital eccentricity and the velocity semi-amplitude very well. Figure\, \ref{correl_ecc_p_K1} shows the covariance between eccentricity, semi-amplitude $K$, and orbital period for 10\,000 Monte Carlo realizations.\\
An increased eccentricity can lead to a longest orbital period with a slightly higher semi-amplitude. However, the lowest limit of these two parameters is quite sharp and a more circular orbit does not change $K$ and $P$. With less than an orbital phase coverage, we can not exclude a long-term trend in the measurements. Adding a positive linear drift to the model leads to a slight decrease in the period (down to a factor of 0.8) but does not change the eccentricity which stays above $e<0.82$. Applying the Akaike information criterion \citep{akaike1974}, which attempts to find the model that best explains the data with a minimum of free parameters, reveals that the single Keplerian model without drift fits our measurements better (Fig.\ref{HD98649_orb}). The residuals around the single planet orbital solution for both data sets ($\sigma_{(O-C)_{C98}}=10.31$ ms$^{-1}$, $\sigma_{(O-C)_{C07}}=5.68$ ms$^{-1}$) agree with the precision of the measurements. With a reduced $\chi^{2}$ of 1.35$\pm$0.27 and a G.o.F = 1.42, the single-planet Keplerian model adequately explains the radial-velocity variation. HD\,98649b is in the top five of the most eccentric planetary orbits and the most eccentric planet known with a period longer than 600 days. In the absence of detection of an outer companion, we cannot invoke the Kozai mechanism \citep{kozai1962} as an explanation for the origin of the high eccentricity of the orbit. With an orbital separation of 10.4 AU at apoastron, leading to an orbital separation up to 250 milliarcsecond, HD\,98649b is an interesting candidate for direct imaging. 

\subsection{A ten Jupiter-mass planet on a ten years orbit around HD\,106515A}
HD\,106515A has been observed with CORALIE at La Silla Observatory since May 1999. Twenty-two Doppler measurements were gathered with CORALIE-98 with a typical signal-to-noise ratio of 47 (per pixel at 550\,nm), leading to a mean measurement uncertainty of 8.4 ms$^{-1}$. Twenty-four additional radial-velocity measurements were obtained with CORALIE-07 with a mean signal-to-noise ratio of 61 leading to a mean measurement uncertainty of 6.3\,ms$^{-1}$. HD\,106515Ab has completed slightly more than one orbital revolution since our first measurement (Fig.\ref{HD106515A_orb}), and the phase coverage near the periastron has been greatly improved by the recently gathered observations. Thus our Monte Carlo realizations for a single-planet Keplerian model point to well constrained orbital elements for HD\,106515Ab. The 1$\sigma$ confidence interval for $K$ is 158.2$\pm$2.6 ms$^{-1}$, which leads to a planetary mass of 9.61$\pm$0.14 M$_{\mathrm{Jup}}$. For the same confidence interval, the eccentricity and period distribution results in the values $e$ = 0.572$\pm$0.011 and $P$ = 3630$\pm$12 days with a semi-major axis of 4.590$\pm$0.010 AU. The orbital separation ranges from 1.96 AU at periastron to 7.22 AU at apoastron which corresponds respectively to 56 and 205 milliarcsecond. The residuals to the fit are somewhat greater than the precision of the measurements ($\sigma_{(O-C)_{C98}}=9.63$ ms$^{-1}$, $\sigma_{(O-C)_{C07}}=6.66$ ms$^{-1}$) and are probably caused by a combination of stellar jitter and light contamination of the second component of the binary. However, with a reduced $\chi^{2}$ of 1.55$\pm$0.28 and a G.o.F = 2.16, the data are well modeled with a single Keplerian orbit. The presence of a stellar companion suggests that the eccentric orbit of HD\,106515Ab may arise from a Kozai mechanism (see Sect.\,\ref{conclusions}).

\subsection{HD\,166724b:  a 14-year period companion of 3.5 Jupiter mass}
We began observing HD\,166724 in September 2001 and have acquired 99 Doppler measurements for this target since then. Eighteen of these were obtained with CORALIE-98 with a signal-to-noise of 27 leading to a mean measurement uncertainty of 10.6 ms$^{-1}$. Twenty-nine more radial velocities were gathered with CORALIE-07 and have a mean final error of 7.3 ms$^{-1}$ with S/N = 51 per pixel at 550 nm. Since HD\,166724 is also part of the HARPS high-precision survey searching for low-mass planets around dwarf stars, fifty-five radial-velocities with higher signal-to-noise and precision have been obtained with this spectrograph since April 2004. The HARPS data on HD\,166724 have a mean measurement uncertainty (including photon noise and calibration errors) of 0.95 ms$^{-1}$ corresponding to a typical signal-to-noise ratio per pixel at 550 nm of 91. To account for the slightly high stellar activity (log\,$R'_{HK}=-4.73$), a velocity jitter of 3.5 ms$^{-1}$ was quadratically added to the mean radial velocity uncertainty.\\
Even though the observing time span of 11 years does not cover the putative complete orbital period, the shape of the orbit is well constrained by the Doppler measurements gathered near and after the passage through periastron, which constrain the eccentricity and the semi-amplitude. However, a second observation at the periastron passage is needed to definitively constrain the orbital period and the other orbital parameters. Running Yorbit followed by MCMC simulations led to a period of 14.1$_{-1.3}^{+1.9}$ years with an eccentricity of $e$ = 0.734$\pm$0.020. A correlation (R=0.7) remains between eccentricity and period (Fig.\ref{correl_ecc_p_K1}), however the value of semi-amplitude ($K$ = 71.0$\pm$1.7 m$^{-1}$) is decorrelated from these two parameters. The corresponding best Keplerian fit and time series radial velocities using C98, C07, and HARPS are plotted in Fig.\ref{HD166724_orb}. The single-planet Keplerian fit yields a reduced $\chi^{2}$ of 1.24$\pm$0.16, however the residuals around the model remain fairly high with an rms\,=\,3.57\,ms$^{-1}$ for HARPS. The large re-emission peak in the CaII region and the relatively high value of log\,$R'_{HK}$ observed for HD\,166724 point toward influence of the stellar jitter to explain this remaining excess of variability. 

\subsection{HD\,196067b:  a Jovian companion (6.9 M$_{\mathrm{Jup}}$) on a ten-year orbit}
Eighty-two observations have been obtained on HD\,196067 since September 1999. Thirty radial velocities were gathered with CORALIE-98 and fifty-two with CORALIE-07. The mean signal-to-noise ratio and measurement uncertainty are 72 and 7.5\,ms$^{-1}$ for C98 and 115 and 5.8\,ms$^{-1}$ for C07. The time series velocities and the corresponding best-fit Keplerian orbit are plotted in Fig.\ref{HD196067_orb}. The rms to the Keplerian fit is 8.88\,ms$^{-1}$ for C98 and 9.92\,ms$^{-1}$ for C07 with G.o.F = 6.55 and a reduced $\chi^{2}$ of 2.48$\pm$0.26. The residuals around the single Keplerian model are slightly more than the precision of the measurements for C07, which points toward the stellar jitter or light contamination and/or guiding errors induced by the second component of the binary. Due to the poor coverage during the closest approach, the data are somewhat overfitted for C98. This explains why the rms to the fit is better for this instrument.\\
The radial velocity of HD\,196067 remained very stable during the first three years of measurement and the observing frequency of the target was diminished. This explains the poor data sampling before the periastron. Because HD\,196067b has completed one orbital revolution since the beginning of the CORALIE survey, we have only gathered two observations near the radial velocity minimum. This therefore allows us to compute the semi-amplitude with a 68.3\% confidence interval of 80 to 256 ms$^{-1}$, yielding to a derived planetary mass of 5.8 to 10.8 M$_{\mathrm{Jup}}$. The period is, however, not perfectly constrained with a minimum value of 9.5 years corresponding to an eccentricity of $e$=0.57. The choice of a higher eccentricity of 0.84 tends to increase the period up to 10.6 years. Finally, the Kozai mechanism induced by the second stellar component of the system can be invoked to explain the eccentricity of the planetary orbit (see Sect. \ref{conclusions}).

\subsection{HD\,219077b:  an ten Jupiter-mass planet on a fifteen-year orbit}
HD\,219077 has been observed with CORALIE at La Silla Observatory since June 1999. Ninety-three Doppler measurements were gathered on this target during the past 13.3 years with C98, C07, and HARPS. The typical signal-to-noise ratio (per pixel at 550\,nm) for each instrument is respectively 46,112, and 211 corresponding to a mean measurement uncertainty of 8.3,5.7,1.0 ms$^{-1}$. As HARPS measurements are contemporary with both CORALIE data sets, the velocity offsets between the spectrographs are therefore well constrained. The radial velocities time series display a linear drift of 13 ms$^{-1}$yr$^{-1}$ over the first eight years of observation. This trend was compatible with a brown dwarf companion at a distance that would make the detection possible with the resolution of the Very Large Telescope. A direct observation of the companion of HD\,219077 was attempted in 2006 by Montagnier et al. (in prep) with the simultaneous differential imaging mode of the adaptive optics instrument NACO. No detection was reported in their high-contrast images, and they can exclude a companion that is more massive than 100 M$_{\mathrm{Jup}}$. Combining their observations and the linear drift in the CORALIE data, they predict that HD\,219077b has a probability of 0.75 to be a brown dwarf.\\
Since 2010, an acceleration in the radial velocity was observed in the time series of CORALIE-07. In September 2011, the cadence of Doppler measurements was increased in order to optimize the data sampling near the periastron passage. Even if the orbital period is still not covered today, this strategy has allowed the orbital parameters to be constrained through MCMC simulations. The C98, C07, and HARPS data sets and the corresponding best Keplerian fit are plotted in Fig.\,\ref{HD219077_orb}. The remaining correlations between eccentricity, period, and semi-amplitude, characterized by the MCMC method, are presented in Fig.\,\ref{correl_ecc_p_K1}. Even if the semi-amplitude is well constrained ($K$ = 181.4$\pm$0.8 m$^{-1}$) and the 1-$\sigma$ confidence interval of the eccentricity is relatively small (0.767 up to 0.773), the corresponding period range remains around 0.7 years ($14.7\,yr \leq P \leq 15.4\,yr$), implying a minimum mass $m_{p}\,$sin$\,i$ for HD\,219077b between 10.30 and 10.48 M$_{\mathrm{Jup}}$.

\subsection{HD\,220689b:  a Jupiter analogue on a six years orbit}
\label{HD220689_section}
The first CORALIE measurement of HD\,220689 was gathered in September 2000. The first eleven of these were obtained with CORALIE-98 with a signal-to-noise of 52 and a mean measurement uncertainty of 8.1 ms$^{-1}$. Thirty-three more radial velocities were gathered with CORALIE-07 and have a mean final error of 5.8 ms$^{-1}$ with S/N = 99 per pixel at 550 nm. A very significant peak with $FAP<10^{-5}$ is identified around 2200 days in the Lomb Scargle periodograme \citep{gilliland1987wls} of the CORALIE time series. This signal corresponds to a Jupiter-mass planet ($m_{p}\,$sin$\,i$= 1.06$\pm$0.09 M$_{\mathrm{Jup}}$) orbiting HD\,220689 with $P$ = 2209$_{-81}^{+103}$ days on a slightly eccentric orbit (e = 0.16$_{-0.07}^{+0.10}$). The residuals to the Keplerian fit are 6.08\,ms$^{-1}$ for the combined data set leading to a reduced $\chi^{2}$ of 1.14$\pm$0.25. The semi-amplitude $K$ = 16.4$\pm$1.5 m$^{-1}$ of HD\,220689b is one of the weakest radial velocities signals unveiled with CORALIE until today. However, the detection of the companion remains very robust with a signal semi-amplitude around three times bigger than the rms to the model. No more significant power is detected in the time series periodograme after subtracting HD\,220689b's orbit.

\section{Updated parameters for three known exoplanets}
\label{updated_param}
\begin{table}
\caption{Observed and inferred stellar parameters for HD\,10647, HD\,30562, and HD\,86226.}
\label{table_stars2} 
\tabcolsep=3.0pt     
\centering          
\begin{tabular}{l l c c c}    
\hline\hline       
Parameters		&	 	&HD10647		&HD30562		&HD86226		\\ 
\hline                    
 Spectral type$^{(a)}$	&		&F9\,V			&F8\,V			&G2\,V			\\  
 $V^{(a)}$		&		&5.52			&5.77			&7.93			\\
 $B-V^{(a)}$		&		&0.551			&0.631			&0.647			\\
 $\pi^{(b)}$		&[mas]		&57.36$\pm$0.25 	&37.84$\pm$0.35		&22.2$\pm$0.8		\\
 $M_{v}$		&		&4.31  			&3.66			&4.66			\\
 $T_{eff}$		&[K]		&6218$\pm$20$^{(d)}$	&5994$\pm$46$^{(c)}$	&5903$\pm$31$^{(c)}$	\\
 log\,$g$		&[cgs]		&4.62$\pm$0.04$^{(d)}$	&4.34$\pm$0.10$^{(c)}$	&4.36$\pm$0.04$^{(c)}$	\\
 $[Fe/H]$		&[dex]		&0.00$\pm$0.01$^{(d)}$	&0.29$\pm$0.06$^{(c)}$	&0.05$\pm$0.03$^{(c)}$	\\
 v$sin\,(i)^{(e)}$ 	&[kms$^{-1} $]	&4.9			&4.1			&2.4			\\
 M$_{*}^{(f)}$		&[M$_{\odot}$]	&1.11$\pm$0.02		&1.23$\pm$0.03		&1.06$\pm$0.03		\\
 $L^{(g)}$		&[L$_{\odot}$]  &1.41			&2.63			&1.06			\\   
 $R^{(f)}$		&[R$_{\odot}$] 	&1.10$\pm$0.02		&1.53$\pm$0.04		&1.02$\pm$0.03		\\   
 log\,$R'_{HK}$		&		&-4.77$\pm$0.01$^{(h)}$	&-5.05$\pm$0.09$^{c)}$	&-4.90$\pm$0.05$^{c)}$	\\
 $P_{rot}^{(i)}$	&[days]		&10$\pm$3		&26$\pm$4		&23$\pm$4		\\
 Age$^{(f)}$		&[Gyr]		&1.4$\pm$0.9		&3.6$\pm$0.5		&1.5$\pm$1.3		\\     
\hline                  
\end{tabular}
\tablefoot{
\tablefoottext{a}{Parameter from HIPPARCOS \citep{esa1997}.}
\tablefoottext{b}{Parallax from the new Hipparcos reduction \citep{vanleeuwen2007}.}
\tablefoottext{c}{Parameter derived using CORALIE spectra.}
\tablefoottext{d}{Parameter derived using HARPS spectra \citep{sousa2008}.}
\tablefoottext{e}{Parameter derived using CORALIE CCF \citep{santos2001feh}.}
\tablefoottext{f}{Parameter derived from \cite{girardi2000} models.}
\tablefoottext{g}{The bolometric correction is computed from \cite{flower1996}.}
\tablefoottext{h}{Parameter derived using HARPS spectra \citep{lovis2011}.}
\tablefoottext{i}{From the calibration of the rotational period vs. activity \citep{mamajek2008}.}
}
\end{table}

   \begin{figure}
   \centering
   \includegraphics[width=9.2cm]{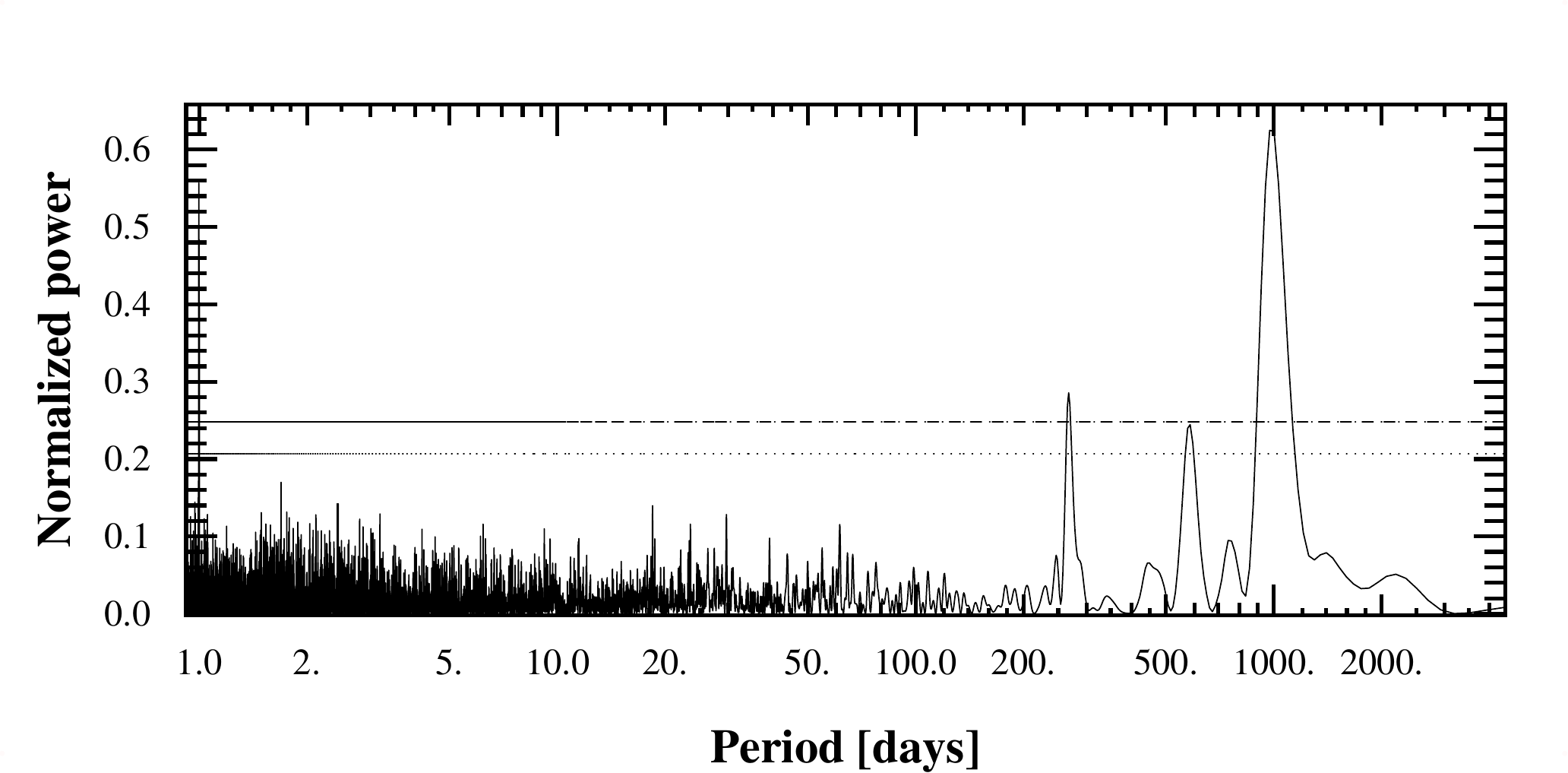}
   \includegraphics[width=9.2cm]{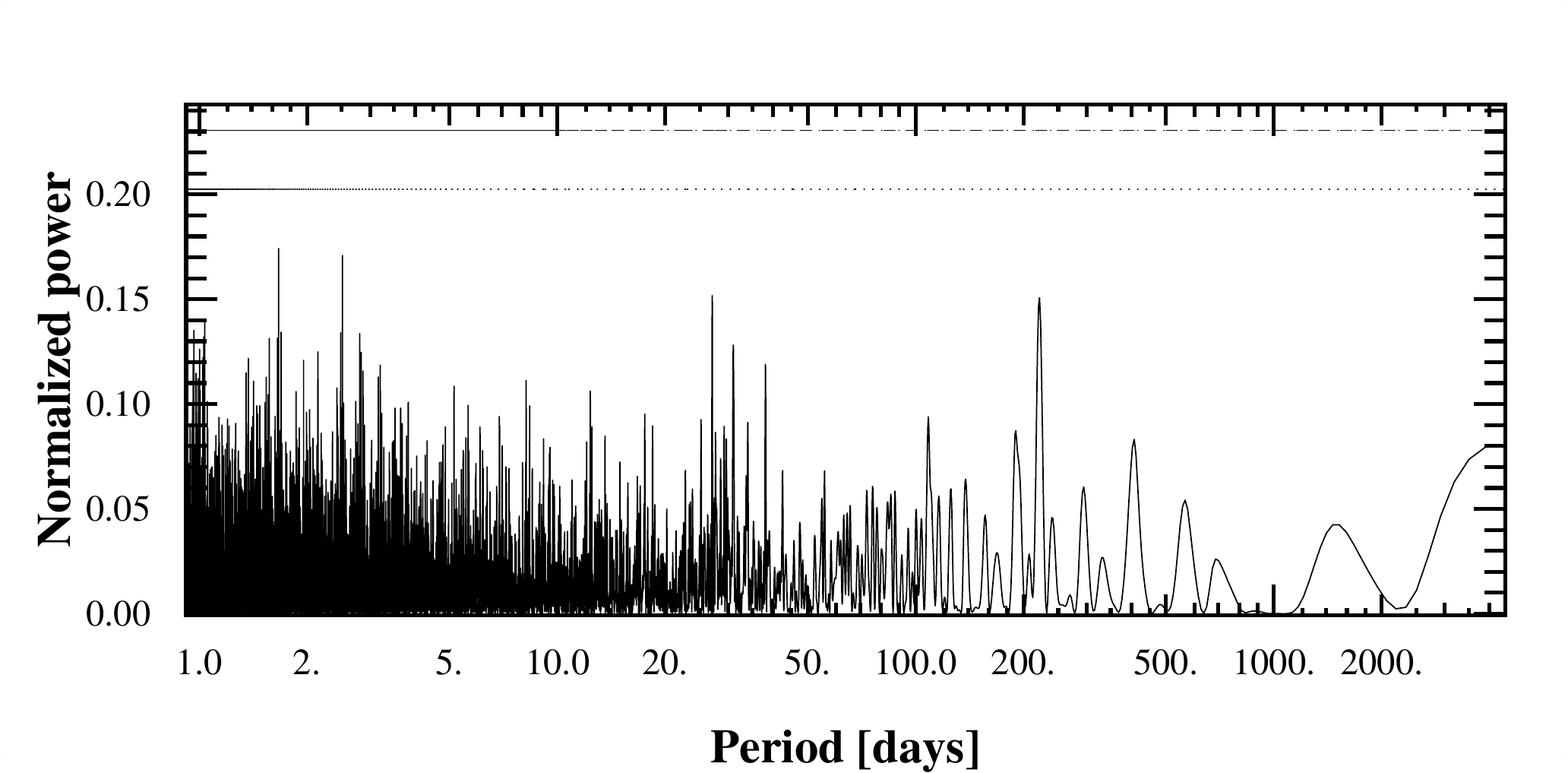}
   \caption{Lomb-Scargle periodogram of the CORALIE radial velocities for HD\,10647 (a). The dotted line indicates the 10$^{-2}$ false alarm probability ($3\sigma$) and the dotted line the corresponding $1\sigma$ limit. The peak at 267 days, is the one-year alias of the planetary signal at P=1000 days. It disappears in the residual's Lomb-Scargle (b), which no longer shows any signal.}
   \label{HD10647_wls_obs}
   \end{figure}
   \begin{figure}
   \centering
   \includegraphics[width=9.2cm]{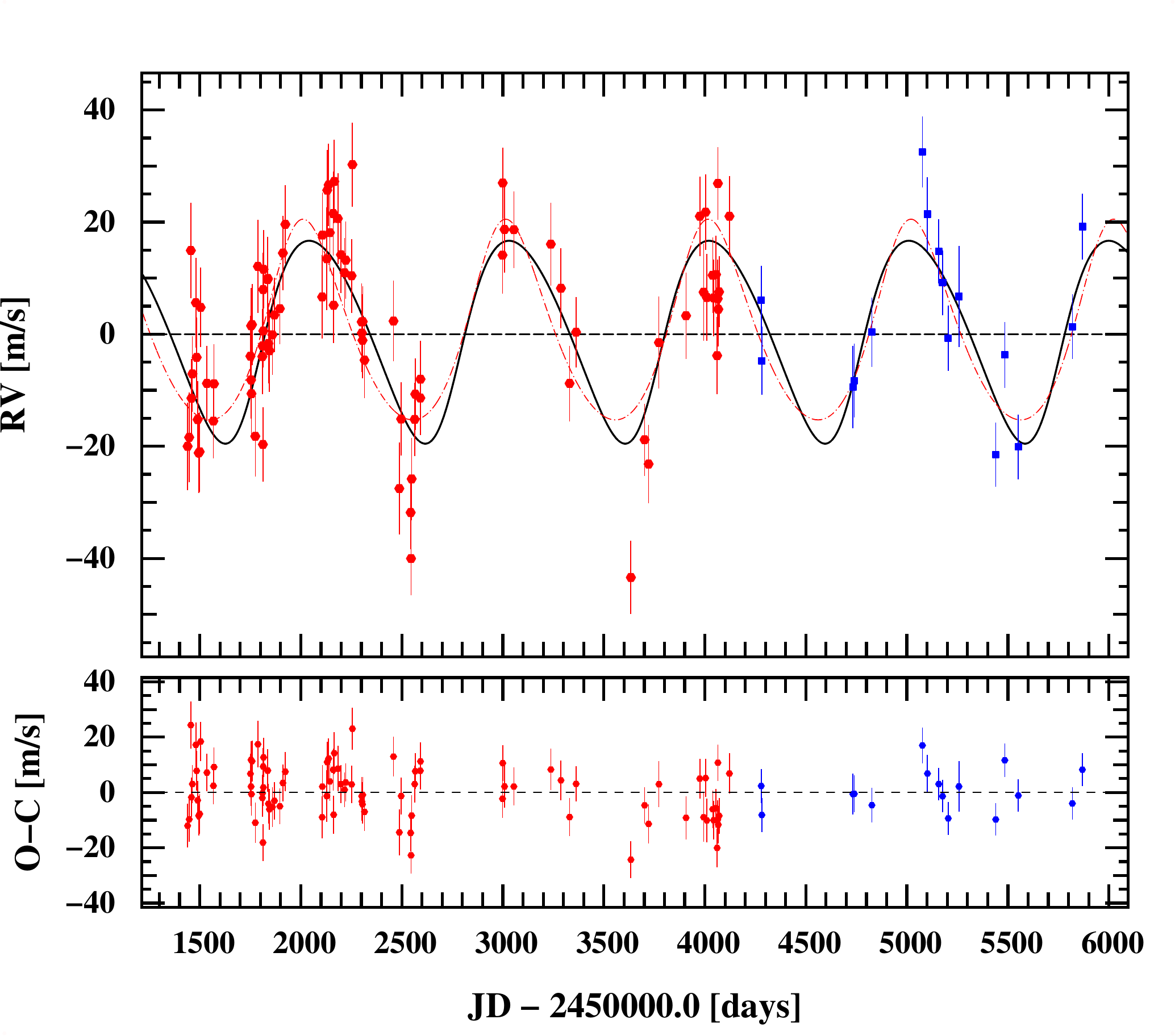}
   \includegraphics[width=9.2cm]{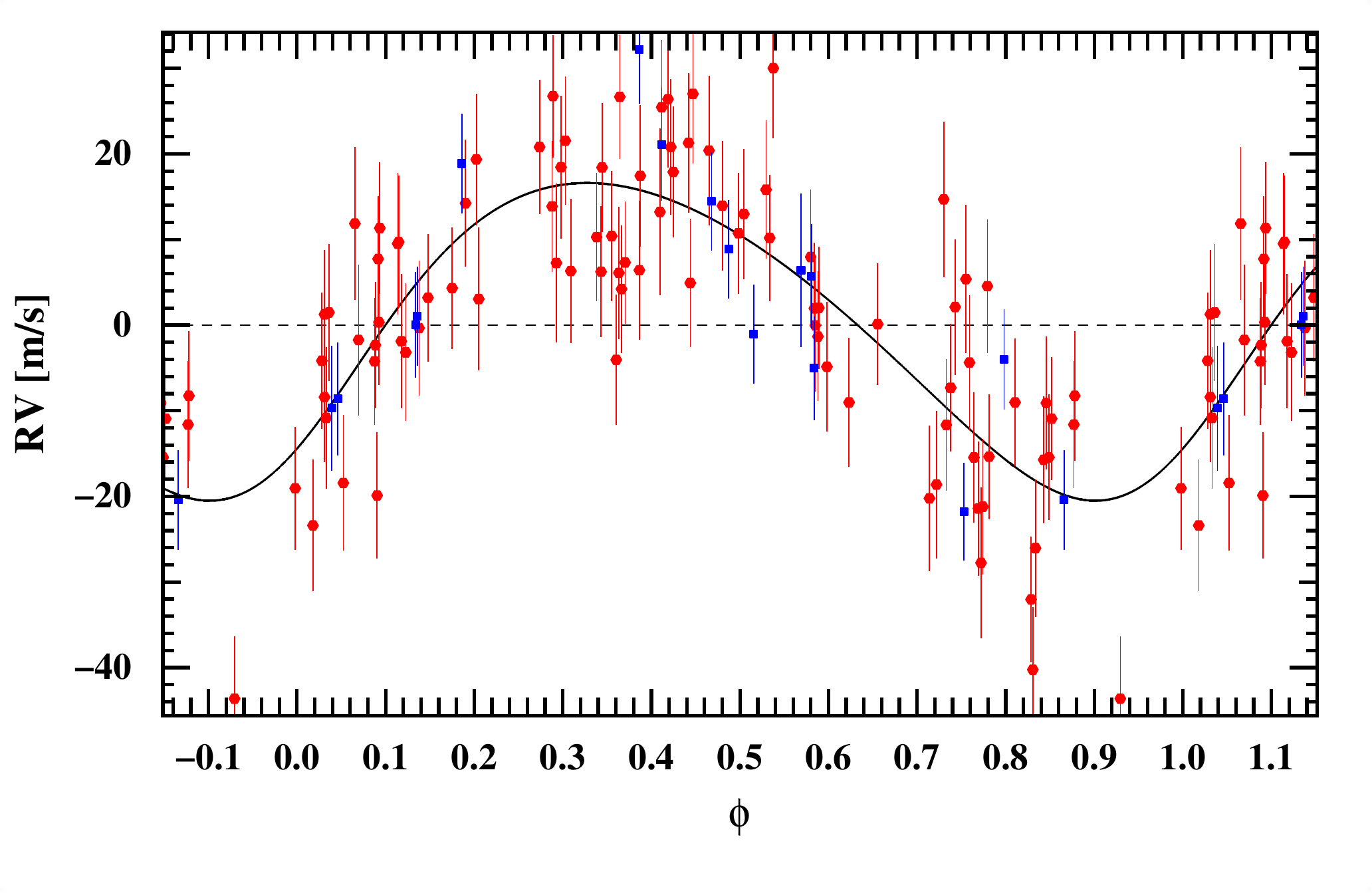}
   \caption{Radial-velocity measurements as a function of Julian Date obtained with CORALIE-98 (red)
      and CORALIE-07 (blue) for HD\,10647. The best single-planet Keplerian model is represented as a black curve while the red dotted line represents the orbital radial velocity signature as expected from the parameters published by \cite{butler2006}. The residuals are displayed at the bottom
      of the top figure and the phase-folded radial-velocity measurements are displayed in the bottom diagram.}
   \label{HD10647_orb}
   \end{figure}
   \begin{figure}
   \centering
   \includegraphics[width=9.2cm]{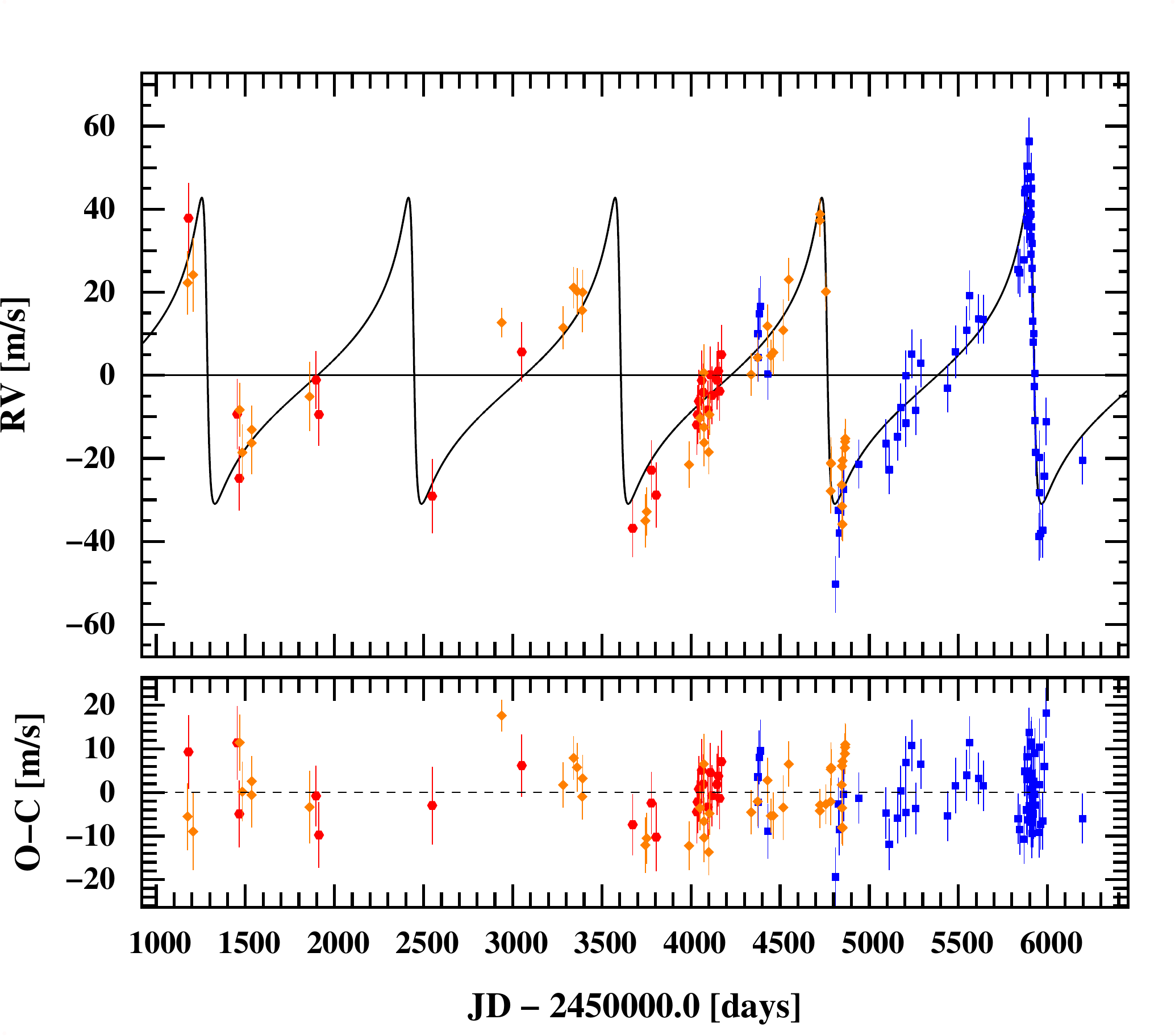}
   \includegraphics[width=9.2cm]{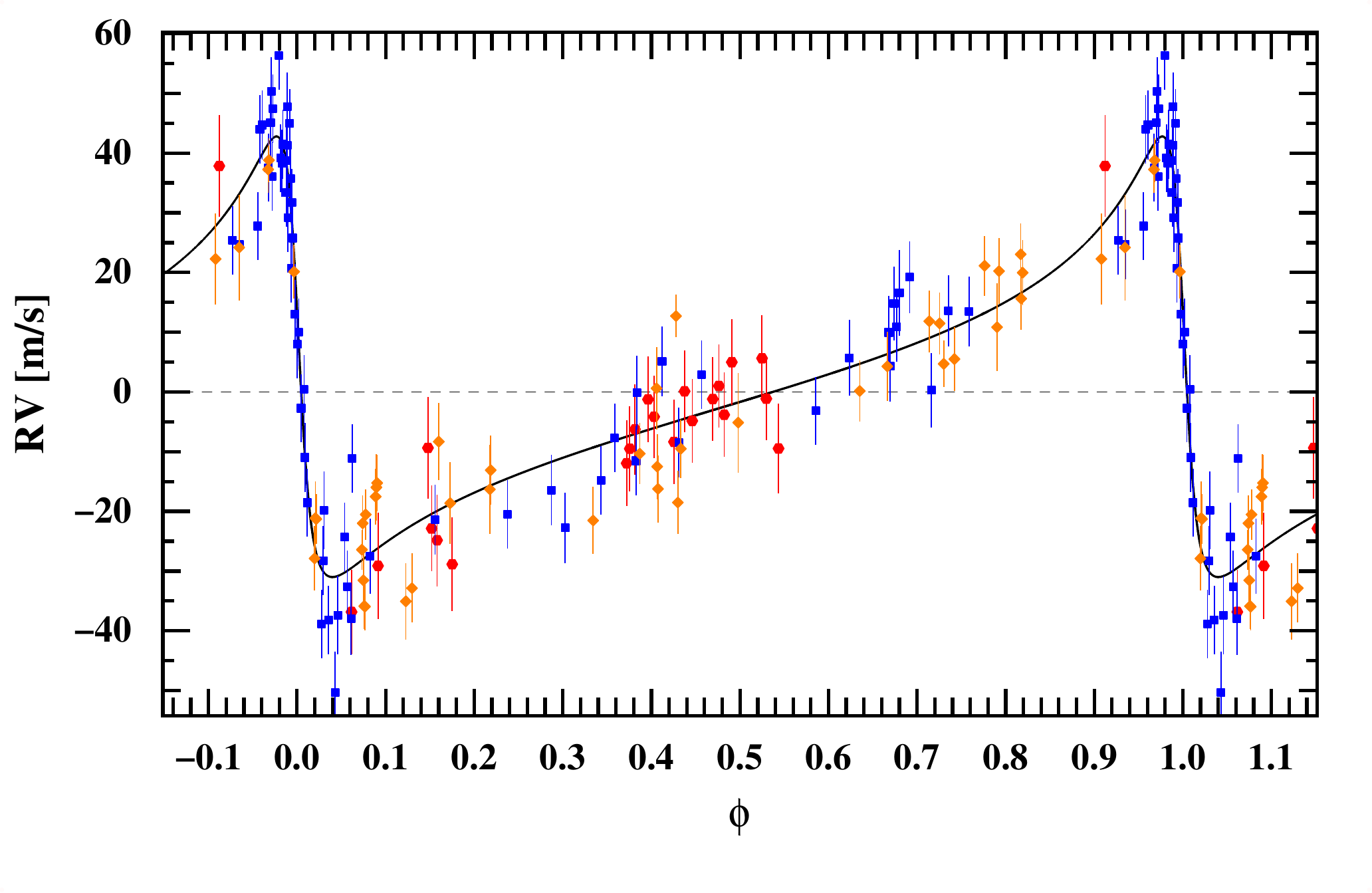}
      \caption{Radial-velocity measurements as a function of Julian Date obtained with CORALIE-98 (red)
      and CORALIE-07 (blue) for HD\,30562. Data from Lick Observatory \cite{fischer2009longp} are plotted in yellow. The best single-planet 
      Keplerian model is represented as a black curve. The residuals are displayed at the bottom
      of the top figure and the phase-folded radial-velocity measurements 
      are displayed in the bottom diagram.}
       \label{HD30562_orb}
   \end{figure}
   \begin{figure}
   \centering
   \includegraphics[width=9.2cm]{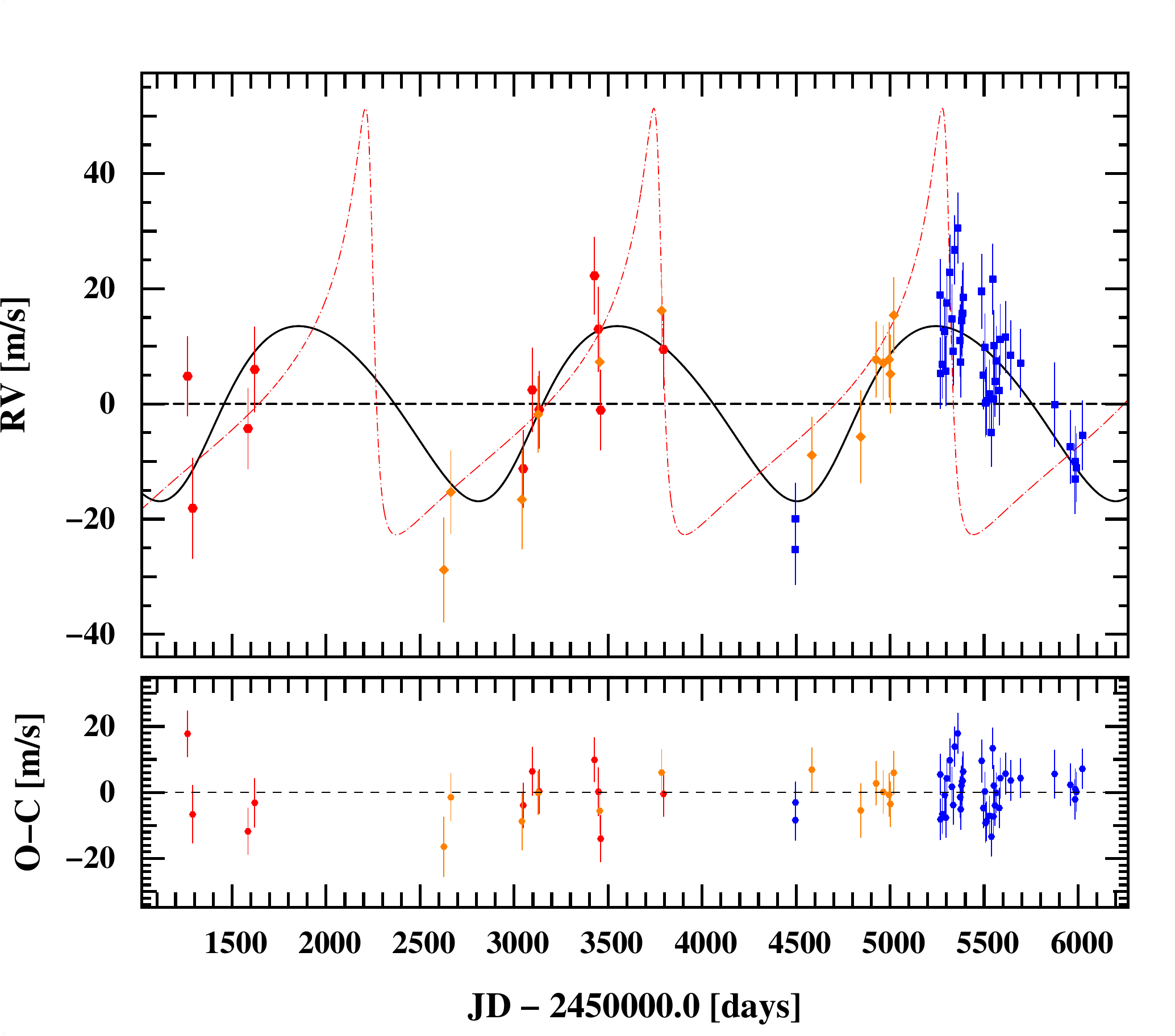}
   \includegraphics[width=9.2cm]{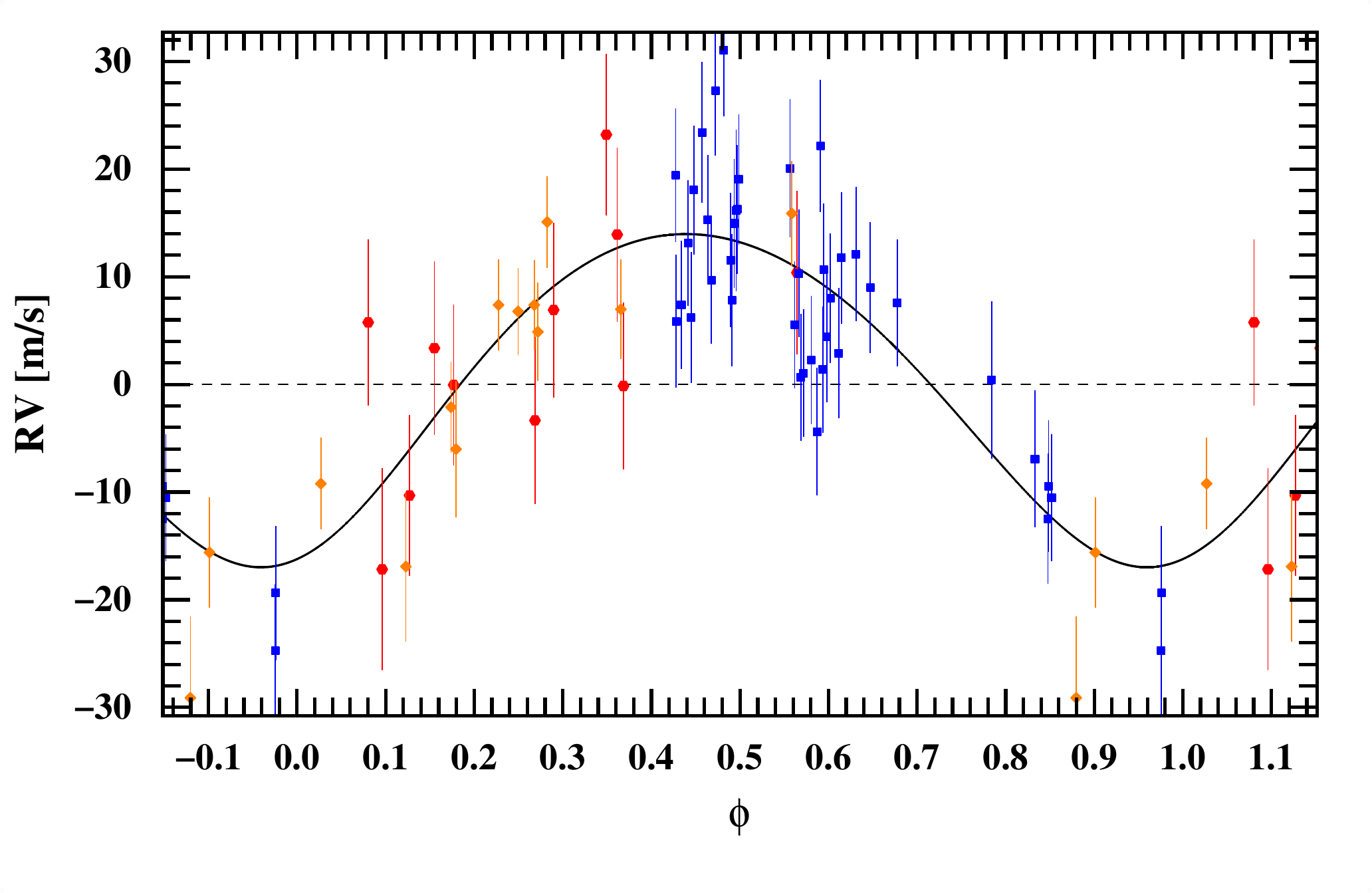}
      \caption{Radial-velocity measurements as a function of Julian Date obtained with CORALIE-98 (red)
      and CORALIE-07 (blue) for HD\,86226. The best single-planet 
      Keplerian model is represented as a black curve while the red dotted line represents the orbital radial velocity signature as expected from the parameters published by \cite{arriagada2009}. The residuals are displayed at the bottom
      of the top figure and the phase-folded radial-velocity measurements 
      are displayed in the bottom diagram.}
       \label{HD86226_orb}
   \end{figure}

\begin{table}
\caption{Single planet Keplerian orbital solutions for HD\,10647, HD\,30562, and HD\,86226.}   
\label{table_orbits2}
\tabcolsep=2.1pt      
\centering      
\begin{tabular}{l l c c c c c c c c}     
\hline\hline       
Parameters		&	 			&HD10647		&HD30562		&HD86226		\\ 
\hline
$P$			&[days]				&989.2$\pm$8.1		&1159.2$\pm$2.8		&1695$\pm$58		\\
$K$			&[m$^{-1} $]			&18.1$\pm$1.7		&36.8$\pm$1.1		&15.3$\pm$1.7		\\
$e$			&				&0.15$\pm$0.08		&0.778$\pm$0.013	&0.15$\pm$0.09		\\
$\omega$		&[deg]				&212$\pm$39		&78.4$\pm$3.0		&193$\pm$61		\\
$T_{0}$			&[JD]				&2453654$\pm$99		&2453601.4$\pm$2.8	&2454821$\pm$288	\\
\hline
$a_{1}\,$sin$\,i$	&[$10^{-3}$ AU]			&1.71$\pm$0.15		&2.47$\pm$0.08		&2.49$\pm$0.28		\\	
$f_{1}(m)$		&[$10^{-9}$ M$_{\odot}$]	&0.70$\pm$0.18		&1.49$\pm$0.15		&0.73$\pm$0.24		\\
$m_{p}\,$sin$\,i$	&[M$_{\mathrm{Jup}}$]		&0.94$\pm$0.08		&1.373$\pm$0.047	&0.92$\pm$0.10		\\
$a$			&[AU]				&2.015$\pm$0.011	&2.315$\pm$0.004	&2.84$\pm$0.06		\\
\hline                    
$\gamma_{C98}$		&[kms$^{-1} $] 			&27.6782		&77.2364		&19.7389		\\  
			&				&$\pm$0.0011		&$\pm$0.0021		&$\pm$0.0039		\\
$\Delta V_{C07-C98}$	&[ms$^{-1} $] 			&-2.4$\pm$2.2		&2.9$\pm$2.4		&-3.0$\pm$4.0		\\
$\Delta V_{^{Lick,}_{MIKE}-C98}$&[ms$^{-1} $] 		&			&-77237.8$\pm$2.3	&-19734.4$\pm$4.1	\\
\hline
$\mathrm{N_{mes}}$ 	&				&108  (92/16)		&130 (22/64/44)		&65  (11/41/13)		\\
$\Delta T$		&[years]			&13.0			&13.7			&13.0			\\
$\chi_{r}^{2}$		&				&1.65$\pm$0.18 		&1.71$\pm$0.17		&1.43$\pm$0.22		\\  
G.o.F			&				&3.90			&4.64			&2.09			\\
$\sigma_{(O-C)}$ 	&[ms$^{-1}$]			&8.92			&6.76			&6.88			\\     
$\sigma_{(O-C)_{C98}}$ 	&[ms$^{-1}$]			&9.38			&4.97			&8.70			\\     
$\sigma_{(O-C)_{C07}}$ 	&[ms$^{-1}$]			&7.51			&6.88			&6.88			\\    
$\sigma_{(O-C)^{Lick,}_{MIKE}}$&[ms$^{-1}$]		&			&7.42			&5.01			\\			
\hline                  
\end{tabular}
\tablefoot{
Confidence intervals are computed for a 68.3\% confidence level after 10\,000 Monte Carlo trials. $C98$ stands for CORALIE-98 and $C07$ for CORALIE-07. $\Delta T$ is the time interval between the first and last measurements, $\chi_{r}^{2}$ is the reduced $\chi^{2}$, G.o.F. is the Goodness of Fit and $\sigma_{(O-C)}$ the weighted rms of the residuals around the derived solution.
}
\end{table}

\subsection{HD\,10647Ab (HIP\,7978b), a Jupiter-mass planet on a thousand-day period}
\label{HD10647_section}
HD\,10647b was first announced in 2003 by Mayor et al. in the XIVth IAP Colloquium. At this time, the small semi-amplitude of the orbit and the relatively high residuals to the Keplerian fit due to stellar activity, pushed the authors to carefully gather more data before publication.
Three years later, \cite{butler2006} published orbital parameters for HD\,10647b based on 28 radial velocities gathered with the UCLES on the Anglo-Australian Telescope. However, the CORALIE radial velocities and parameters were never published before this paper.
HD\,10647 is a bright F9 dwarf located in the Eridanus constellation at $17.43\pm0.08$ pc from the Sun. The star properties are summarized in Table\,\ref{table_stars2}.
Even though no clear prominent re-emission feature in Ca II absorption region can be seen in the HD\,10647 spectra (Fig.\,\ref{caII}), the log\,$R'_{HK}$ value of $-4.77\pm0.01$ points to moderate chromospherical activity. The possibly induced stellar spots combined with a slightly high measured projected rotational velocity of v$sin\,(i)=4.9$ kms$^{-1}$ led to expecting some radial-velocity jitter on a typical timescale on the order of the rotational period ($P_{rot}=10\pm3$ days).
A clear signature around P=1000\,days (Fig.\ref{HD10647_wls_obs}.a) is identified in the Lomb Scargle periodogram of the 108 Doppler measurements gathered with CORALIE since September 1999. This signal corresponds to a Jupiter-mass planet ($m_{p}\,$sin$\,i$ = 0.94$\pm$0.08 M$_{\mathrm{Jup}}$) on a 989.2$\pm$8.1 days orbit with a semi-major axis of a=2.015$\pm$0.011\,AU.
Figure\,\ref{HD10647_orb} shows the combined C98 and C07 times series with the corresponding best-fit Keplerian model and the orbital radial velocity signature, as expected from the parameters published by \cite{butler2006}. The resulting orbital elements for HD\,10647b are listed in Table\,\ref{table_orbits2}. The residuals to the adjusted single-planet Keplerian model show a level of variation that is slightly larger than expected from the precision of the measurements. This points toward the influence of the stellar jitter since the periodogram of the residuals (Fig.\ref{HD10647_wls_obs}.b) no longer shows any signal with significant power. Additionally, a constraint on the radial-velocity solution can be provided by the intermediate astrometric data of the new Hipparcos reduction \citep{vanleeuwen2007}. With the 123 measurements gathered by Hipparcos over 1128 days (1.14 orbits), we applied the method described by \cite{sahlmann2011} to search for the astrometric signature. Hipparcos can constrain the maximum mass of the companion of HD\,10647 to 12.33 $M_{Jup}$ and thus exclude a brown dwarf or M dwarf companion.

\subsection{Independent confirmation and refined orbital parameters of the long-period and eccentric Jovian planet HD\,30562b}
HD\,30562  is identified as a bright F8V in the Hipparcos catalog with an apparent magnitude of $V= 5.77$ at a distance of $26.43\pm0.24$ pc from the Sun. The spectroscopic analysis of the CORALIE spectra results in an effective temperature $T_{eff}=5994\pm46$ and metallicity of $[Fe/H]=0.29\pm0.06$. For this star we derived a mass M$_{*}=1.23\pm0.03$ M$_{\odot}$ with an age of $3.6\pm0.5$ Gyr. \cite{hall2007lprhk} find that HD\,30562 is chromospherically inactive with an activity index of log\,$R'_{HK}=-5.15$ in agreement with the value of log\,$R'_{HK}=-5.05\pm0.09$ derived in the present paper from CORALIE spectra.
We began observing HD\,30562 in January 1999 and have acquired 130 Doppler measurements for this target since then. The first twenty-two of these were obtained with CORALIE-98 with a signal-to-noise of 68 leading and a mean measurement uncertainty of 7.1 ms$^{-1}$. Sixty-four more radial velocities were gathered with CORALIE-07 and have a mean final error of 5.8 ms$^{-1}$ with S/N = 130 per pixel at 550 nm.
In August 2009, HD\,30562b was announced by \cite{fischer2009longp} based on forty-four data points gathered at the Lick Observatory. Since a good phase coverage is needed to constrained the high eccentricity of the orbit, we decided to combine the three data sets to refine the orbital parameters. The corresponding best Keplerian fit and time series radial velocities from C98, C07, and Lick are plotted in Fig.\,\ref{HD30562_orb}. Since each instrument has its own velocity zero point, the velocity offsets between the three spectrographs are additional free parameters in the model. The data sets have contemporary observations, so the velocity offsets are well constrained. The residuals to the adjusted single-planet Keplerian model show a level of variation of $\sigma$=6.76 ms$^{-1}$, yielding a reduced $\chi^{2}$ of 1.71$\pm$0.17. The rms to the Keplerian fit is 4.97\,ms$^{-1}$ for CORALIE-98, 6.88\,ms$^{-1}$ for CORALIE-07, and 7.42\,ms$^{-1}$ for Lick data. The resulting orbital parameters are $P$ = 1159.2$\pm$2.8 days, $e$ = 0.778$\pm$0.013, $K$ = 36.8$\pm$1.1 m$^{-1}$, implying a minimum mass $m_{p}\,$sin$\,i$ = 1.373$\pm$0.047 M$_{\mathrm{Jup}}$ orbiting with a semi-major axis $a$=2.315$\pm$0.004 AU. These results are consistent with the parameters by \cite{fischer2009longp}. The combined fitting of the data sets leads to a slightly higher eccentricity and semi-amplitude, implying a planetary mass around 5\% above the value previously announced. Interestingly, due to the high eccentricity, the transit probability of HD\,30562b is quite high ($\approx$1\%) for a planet with a revolution period in this range.

\subsection{Independent confirmation and new orbital parameters of the long-period Jupiter-mass planet around HD\,86226}
HD\,86226 has a spectral type G2V with an astrometric parallax of  $\pi = 22.2\pm0.8$\,mas, which sets the star at a distance of $45.0\pm1.6$ pc from the Sun in the Hydra constellation. The star is chromospherically quiet with a log\,$R'_{HK}$ of -4.90 $\pm$ 0.05 leading to a stellar rotation period of $P_{rot}=23\pm4$ days. The first measurement of HD\,86226 was gathered in March 1999. Thirteen observations were obtained with CORALIE before the announcement of HD\,86226b made by \cite{arriagada2009} based on thirteen MIKE measurements. At this time, despite a small excess of variability, no significant signal could be unveiled in the CORALIE data. Since March 2010 a new observation campaign has been carried out on HD\,86226, leading to thirty-nine new Doppler measurements. The observing frequency was relatively high in order to observe the fast radial velocity decrease predicted by published orbital parameters. Figure\,\ref{HD86226_orb} shows the combined C98, C07, and MIKE times series with the corresponding best-fit Keplerian model and the orbital radial velocity signature, as expected from the parameters published by \cite{arriagada2009}. The new CORALIE observations confirm the presence of a Jovian planet ($m_{p}\,$sin$\,i$ = 0.92$\pm$0.10) on a 4.6-year period around HD\,86226. However, the combined fitting of the data sets leads to an orbit eccentricity and a much lower semi-amplitude than previously announced. The resulting orbital elements for HD\,86226b are listed in Table\,\ref{table_orbits2}. The case of HD\,86226b is a good illustration of the bias toward overestimated $M_{p}\,$sin$\,i$ and eccentricity for a moderate number of observations and relatively low signal-to-noise ratio of $K/\sigma \lesssim 3$ \citep{shen2008_ecc}.

\begin{figure}
   \centering
   \includegraphics[width=8.8cm]{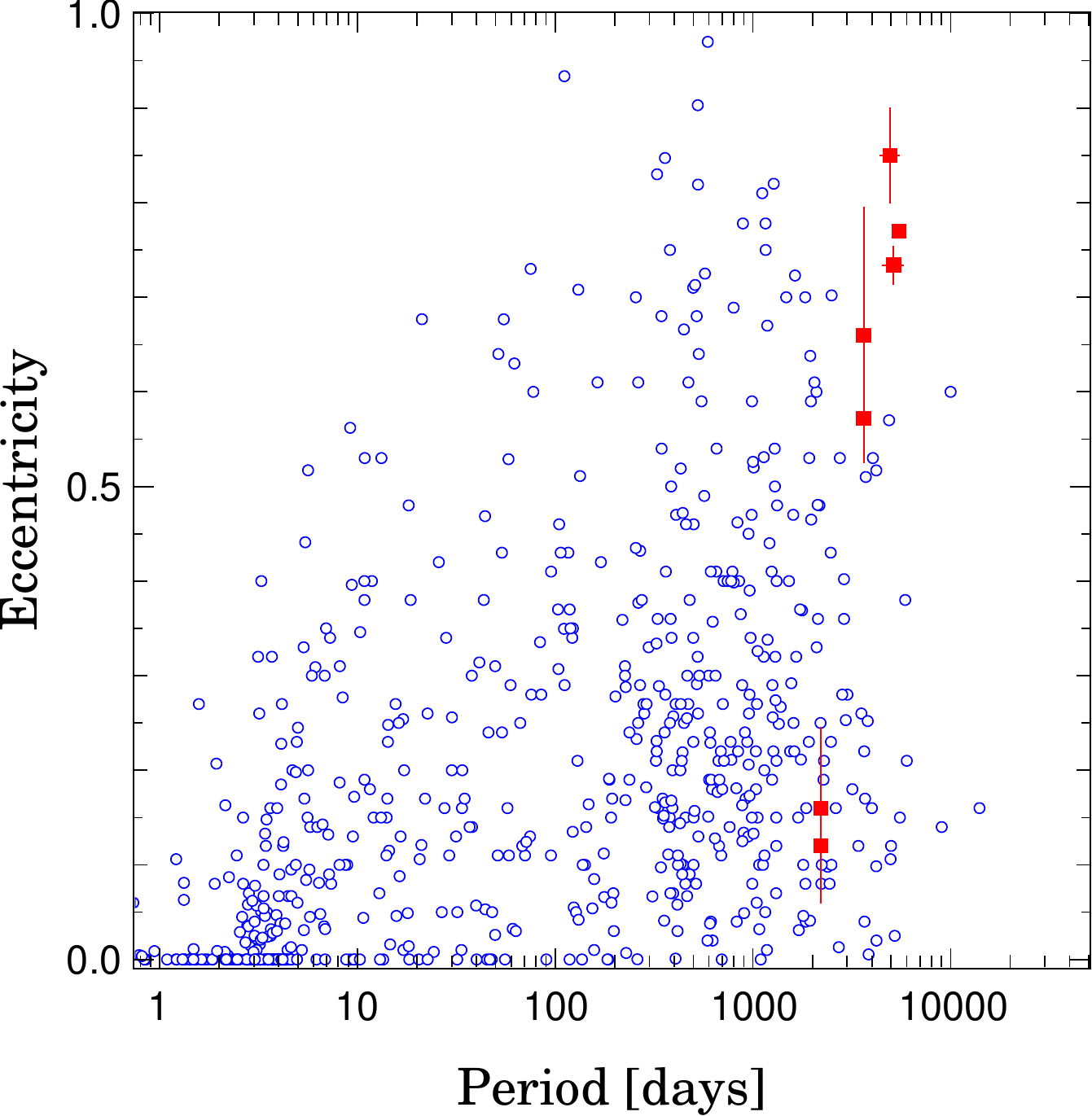}
   \caption{Observed orbital eccentricity vs. period for the known extra-solar planets from the exoplanet.eu database \citep{schneider2011}. The parameters of the seven companions unveiled in the present paper are plotted in red.}
   \label{eccvsp}
\end{figure}

\section{Discussion \& conclusions}
\label{conclusions}
We have reported in this paper the discovery of seven giant planets with long orbiting periods discovered with the CORALIE echelle spectrograph mounted on the 1.2-m Euler Swiss telescope located at La Silla Observatory. Interestingly, we notice that the hosts of these long-period planets show no metallicity excess. Five of them - HD\,98649b, HD\,106515Ab, HD\,166724b, HD\,196067b, and HD\,219077b - have periods around or above ten years, and they constitute a substantial addition to the previously known number of planets with periods in this range. These five exoplanets have surprisingly high eccentricities above $e>0.57$ (Fig.\,\ref{eccvsp}). HD\,98649b, HD\,166724b, and HD\,219077b ($e=0.85$, $e=0.734$, and $e=0.770$) are even the three most eccentric planets with a period longer than five years. Our Monte Carlo and MCMC simulations show that this parameter is well constrained for these targets and that no bias toward overestimated eccentricities \citep{shen2008_ecc} could be invoked. We notice that, due to their high eccentricity orbits, the orbital separation range up to  10.4 and 11.0 AU at apoastron for HD\,98649b and HD\,219077b. This can lead to an angular separation up to 250 and 375 milliarcsecond, which makes two interesting candidates for direct imaging. In both host stars HD\,106515A and HD\,196067, a third massive body is present in the system. To investigate whether the eccentricity of the orbit may arise from the Kozai mechanism \citep{kozai1962}, we estimate their Kozai period of eccentricity oscillation \citep{mazeh1979,ford2000kozai} 
\begin{equation}
\label{pkozai}
P_{Kozai}\approx P_{planet} \left(\frac{M_{p}}{M_{s}} \right) \left(\frac{a_{b}}{a} \right)^3 \left(1-e_{b}^{2}\right)^{3/2} 
\end{equation}
where $a$ and $P_{planet}$ are the semi-major axis and the period of the planetary orbit, $M_{p}$ and $M_{s}$ the mass of the primary and the secondary stars, $a_{b}$ and $e_{b}$ the semi-major axis and the eccentricity of the binary. Assuming as eccentricity distribution $f(e)=2e$ for the binaries with $P \geqslant 1000$ days \citep{duquennoy1991} and taking the median of the distribution ($e_{b}=\frac{1}{\sqrt{2}}$), we obtain $P_{Kozai} = 3.3 \times 10^{6}$ years for HD\,106515Ab and $P_{Kozai} = 3.5 \times 10^{7}$ years for HD\,196067. The period of eccentricity oscillations $P_{Kozai}$ has to be compared to age of the system and to the apsidal precession due to the relativistic correction to the Newtonian equation of motion \citep{holman1997}, which is given by
\begin{equation}
\label{prel}
P_{rel} = P_{planet} \frac{c^{2}\left(1-e^{2}\right)a}{3 GM_{star}}.
\end{equation}
This yields $P_{rel} = 1.1 \times 10^{9}$ years for HD\,106515Ab and $P_{rel} = 7.4 \times 10^{8}$ years for HD\,196067b. For the two planets the eccentricity oscillation half-period is shorter than the age of the system and at least one order faster than the precession due to the relativistic effects. Since both binary stars have semi-major axis wider than 40 AU, we can assume that the planet and the binary orbital planes have uncorrelated inclinations \citep{hale1994}. Thus the Kozai pumping mechanism is a viable explanation for the eccentricities of the HD\,106515Ab and HD\,196067b orbits. Because the presence of a third massive body cannot be inferred from the data of HD\,98649b, HD\,166724b, and HD\,219077b, the origin of the eccentricity of these systems remains unknown.

\begin{figure}
   \centering
   \begin{minipage}[]{.49\linewidth}
    \begin{center}
       \includegraphics[width=4.5cm]{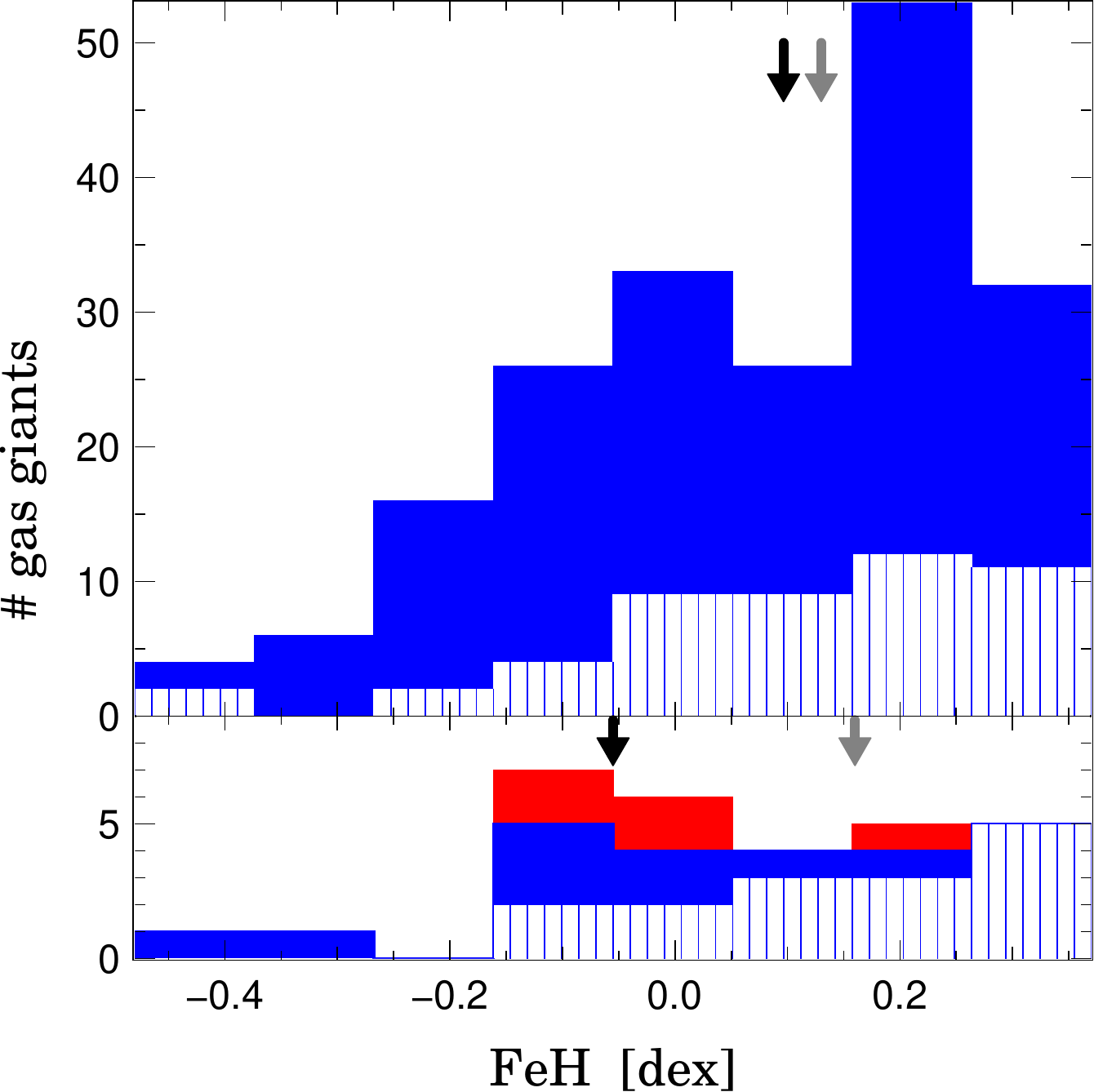}
    \end{center}
   \end{minipage}
   \hfill
   \begin{minipage}[]{.49\linewidth}
    \begin{center}
       \includegraphics[width=4.5cm]{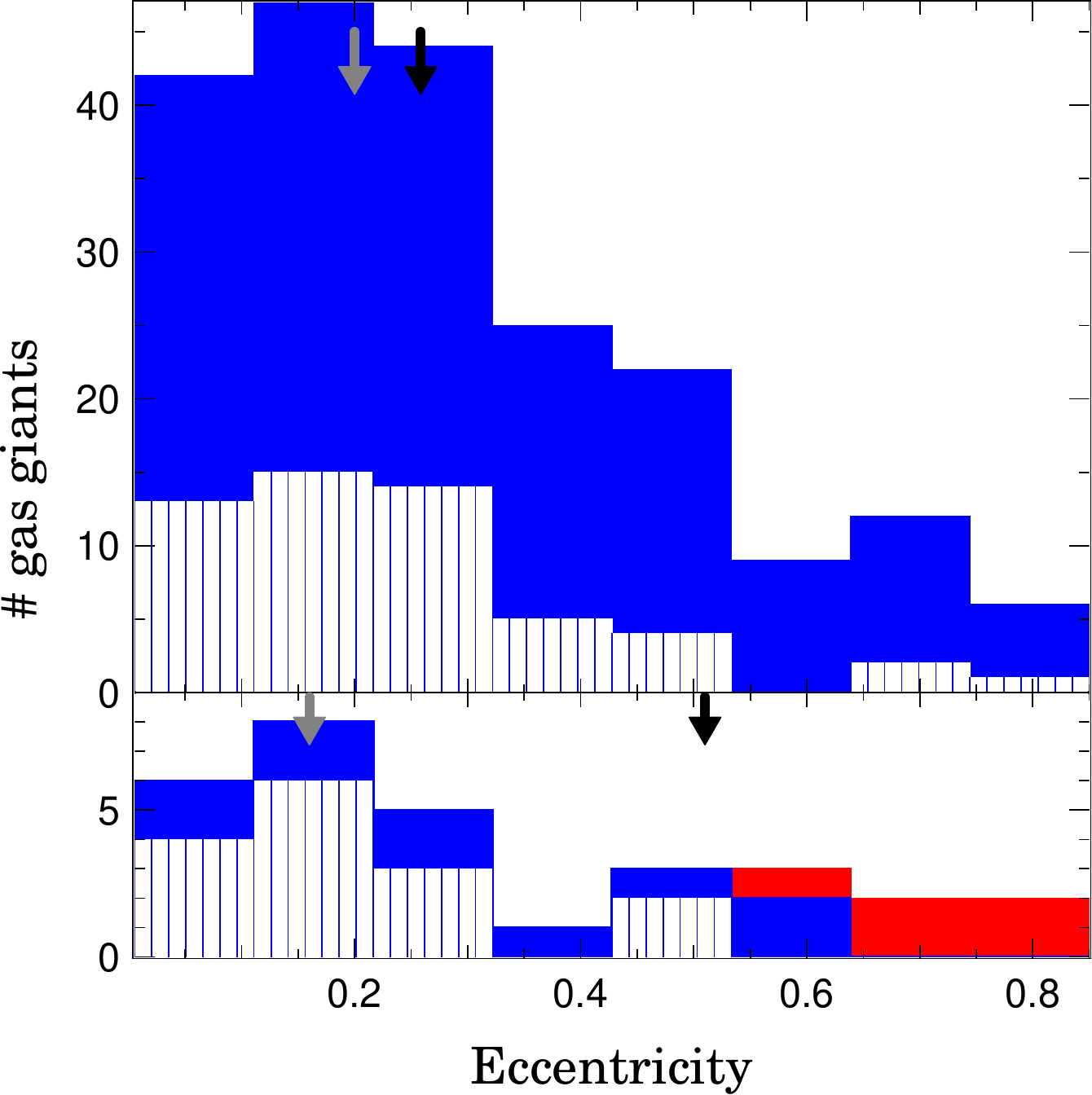}
    \end{center}
   \end{minipage}
   \caption{Distribution of the host star metallicities (left) and orbital eccentricities (right) for the gas giant planets ($>$100 M$_{\oplus}$). The upper histograms represent the planets with semi-major axis between 1 and 4\,AU and the lower quadrant is for planets beyond 4\,AU. The hatched areas plot the contributions of the multiple systems and the red color is used for the 5 companions (with $a>4$\,AU) unveiled in the present paper. The gray arrows mark in every quadrant the median of the distribution for the multiple systems and the black ones are for single-planet systems.}
   \label{histo_feh_ecc}
   \end{figure}

\noindent
Interestingly, half of the known planets with a semi-major axis larger than 4 AU (15/30) are part of multiple systems, whereas this ratio is much lower ($\sim$25\%) for the companions in the 1-4 AU range. The most likely explanation for this is a detection bias induced by the follow-up of the first detection of a smaller period planet, which triggers a follow-up of the star and tends to increase the span of the time series. For many configurations of multiple systems, high eccentricities are incompatible with stability, so these systems are uncommon. As a consequence, the small number of single companions beyond 4 AU biased the eccentricity distribution toward low eccentricities (Fig.\,\ref{histo_feh_ecc}).\\
This trend is probably reinforced when one considers that very eccentric planets are easily overlooked. If one observes the flat part of an eccentric orbit one may detect only small variations, and many surveys would cast the star aside as uninteresting. This point and the large amount of data gathered by the CORALIE survey do not explain the high fraction of very eccentric planets unveiled in the present paper. However, these are small number statistics, and a Kolmogorov-Smirnov test \citep{massey1951} applied to the eccentricity distribution in the range 1-4 AU, and beyond, gives a good probability ($\sim$50\%) for the null hypothesis (i.e. that the samples are drawn from the same distribution) with or without the planets presented in this paper.

\noindent
The same remark applies to the lack of metallicity excess of the hosting stars (of this paper). Even though \cite{boisse2012} has noticed that beyond 4 AU, the well described greater occurrence of gas giants around metal-rich stars \citep{santos2004, fischer2005, mayor2011} seems to remain true, the current statistic is still very poor.  Interestingly, we notice in Fig.\,\ref{histo_feh_ecc} that this correlation appears to be clearer (below and beyond 4 AU) for multiple systems than for single ones. However, an unbiased sample and a statistical approach taking account of the detection efficiency for each star would be needed to confirm this hypothesis.

\noindent
In addition to the seven new companions presented in this paper, we also have published refined orbits for three already known exoplanets - HD\,10647b \citep{butler2006}, HD\,30562b \citep{fischer2009longp}, and HD\,86226b \citep{arriagada2009}. CORALIE time series combined with the published data sets, independently confirms the existence of the three companions. In the case of HD\,86226b, our derived orbital parameters are, however, far from the published ones with an eccentricity decreasing from $e=0.73$ to $e=0.15$ and period more than hundred days longer.

\begin{acknowledgements}
      We are grateful to the Geneva Observatory's technical staff, in
      particular, to L. Weber for maintaining the 1.2-m Euler Swiss telescope and
      the CORALIE Echelle spectrograph. We thank the Swiss National Research
      Foundation (FNRS) and Geneva University for their continuous support
      to our planet search programs. N.C.S. and P.F. would like to thank the European 
      Research Council/European Community for support by under the FP7 through Starting 
      Grant agreement number 239953. N.C.S and P.F. also acknowledge the support from
      Funda\c{c}\~{a}o para a Ci\^{e}ncia e a Tecnologia (FCT) through program Ci\^{e}ncia\,2007
      funded by FCT/MCTES (Portugal) and POPH/FSE (EC) in the form of the grants
      PTDC/CTE-AST/098528/2008 and PTDC/CTE-AST/098604/2008. This research made use 
      of the VizieR catalog access tool operated at the CDS, France. S.Alves acknowledges the CAPES 
      Brazilian Agency for a PDEE fellowship (BEX-1994/09-3).
\end{acknowledgements}
 
\bibliographystyle{aa}
\bibliography{biblio}
\end{document}